\documentclass[11pt,a4paper]{article}

\PassOptionsToPackage{linktocpage,pagebackref}{hyperref}

\usepackage{jheppub}
\makeatletter
\gdef\@fpheader{}
\makeatother

\usepackage[english]{babel}
\usepackage[babel]{csquotes}
\usepackage{mathtools}
\usepackage{float}
\usepackage{braket}
\usepackage{booktabs}
\usepackage{longtable}
\usepackage{nicematrix}
\usepackage{subcaption}
\usepackage{enumitem}
\usepackage[font=small,format=hang,labelfont={sf,bf},figurewithin=section,tablewithin=section]{caption}

\definecolor{darkblue}{rgb}{0.1,0.1,0.7}
\hypersetup{colorlinks,
           linkcolor={darkblue},
           citecolor={darkblue},
           urlcolor={darkblue}
}

\newcommand{\eps}{{\varepsilon}}

\renewcommand{\geq}{\geqslant}
\renewcommand{\leq}{\leqslant}
\newcommand{\cF}{\mathcal{F}}
\newcommand{\cG}{\mathcal{G}}
\newcommand{\cO}{\mathcal{O}}
\newcommand{\cP}{\mathcal{P}}

\newcommand{\R}{\mathbb{R}}
\newcommand{\Z}{\mathbb{Z}}
\newcommand{\ped}[1]{\textormath{\textsubscript{#1}}{_{\mathrm{#1}}}}

\newcommand{\id}{\text{\usefont{U}{bbold}{m}{n}1}}

\newcommand{\ec}{\,,}
\newcommand{\ed}{\,.}
\newcommand{\z}{z}
\newcommand{\zb}{\bar{z}}
\newcommand{\norm}[1]{\left\lVert#1\right\rVert}
\newcommand{\abs}[1]{\left\lvert#1\right\rvert}

\newcommand{\hypgeo}[2]{%
  \operatorname{%
    {\vphantom{\mathnormal{F}}}_{#1}%
    \kern-\scriptspace
    \mathnormal{F}_{#2}%
  }%
}

\let\originalleft\left
\let\originalright\right
\renewcommand{\left}{\mathopen{}\mathclose\bgroup\originalleft}
\renewcommand{\right}{\aftergroup\egroup\originalright}

\renewcommand*{\backref}[1]{}
\renewcommand*{\backrefalt}[4]{\ifcase #1 \or {\scriptsize\,[p.~#2]} \else {\scriptsize\,[pp.~#2]} \fi}

\makeatletter
\patchcmd\NAT@citexnum{\let\NAT@last@num\NAT@num}{\MakeLinkTarget[cite]{}\Hy@backout{\@citeb\@extra@b@citeb}\let\NAT@last@num\NAT@num}{}{\fail}
\makeatother

\title{The Virasoro Bootstrap: from the $c\sim1$ desert to coupled Potts}

\author[a]{Stefanos R.\ Kousvos,}
\author[b,c,d]{Alessandro Piazza,}
\author[e]{Alessandro Vichi}

\affiliation[a]{Department of Physics, University of Torino and INFN, Section of Torino,\\[-1pt] Via P.\ Giuria 1, 10125 Torino, Italy}
\affiliation[b]{SISSA, Via Bonomea 265, I-34136, Trieste, Italy}
\affiliation[c]{INFN, Sezione di Trieste, Via Valerio 2, I-34127, Trieste, Italy}
\affiliation[d]{Abdus Salam Centre for Theoretical Physics, Imperial College London, London SW7 2AZ, UK}
\affiliation[e]{Department of Physics, University of Pisa and INFN, section of Pisa. Largo Pontecorvo 3, I-56127 Pisa, Italy}

\emailAdd{stefanosrobert.kousvos@unito.it} 
\emailAdd{alessandro.piazza25@imperial.ac.uk} 
\emailAdd{alessandro.vichi@unipi.it}

\abstract{We develop a numerical conformal bootstrap for two-dimensional CFTs with $c>1$ based on Virasoro conformal blocks, computed with Zamolodchikov's recursion and optimal polynomial interpolation. As a first test we isolate an island around three decoupled Ising models. We then consider three 3-state Potts models coupled through their energy operators, a candidate irrational CFT. Mixed correlators isolate the decoupled theory in an island, and different gap assumptions, motivated by perturbation theory, yield a peninsula that contains the coupled fixed point but not the decoupled one. Finally, we explore the standard single-correlator bound on the lowest-dimension scalar primary in the OPE of two identical scalars. We find that for central charges just above $c=1$ the bound collapses over a wide range of external dimensions. This excludes any theory whose lightest scalar is absent from its own OPE and lies in this range, and supports the conjecture that no compact CFT exists in this region. We explain several features of these bounds analytically.}

\begin{document}

\maketitle

\section{Introduction}

Conformal field theories (CFTs) in two dimensions occupy a special place among quantum field theories. The idea that critical phenomena are conformally invariant, and the fact that consistency of the operator product expansion (OPE) with crossing symmetry can be used to constrain them, the conformal bootstrap, goes back to the early 1970s~\cite{Polyakov:1970xd,Polyakov:1974gs} (see~\cite{Rychkov:2025zks} for a historical account). The breakthrough came in two dimensions with the work of Belavin, Polyakov and Zamolodchikov~\cite{Belavin:1984vu}, who realised that the local conformal symmetry of a $2d$ CFT is generated by two copies of the infinite-dimensional Virasoro algebra with central charge $c$. Each Virasoro primary organises infinitely many global $\mathrm{SL}(2,\R)$ primaries into a single representation, and correlation functions decompose into Virasoro conformal blocks, which are completely fixed by symmetry in terms of $c$ and the conformal weights. For special values of the weights, the Virasoro representations contain null states, the corresponding correlators obey differential equations, and crossing can be solved exactly~\cite{Belavin:1984vu,Dotsenko:1984ad}. This gave rise to the minimal models: unitarity restricts $c<1$ to a discrete series of values~\cite{Friedan:1983xq}, the simplest of which is the critical Ising model at $c=1/2$. Modular invariance of the torus partition function~\cite{Cardy:1986ie} further constrains the allowed operator content (see~\cite{DiFrancesco:1997nk} for a textbook account).

Minimal models and, more generally, rational CFTs are theories in which crossing is solved with finitely many primaries (under the Virasoro or extended chiral algebras). For $c>1$, however, unitary CFTs are expected to be generically irrational, with infinitely many Virasoro primaries, and no classification is available. Explicit examples with only Virasoro chiral symmetry are nevertheless hard to find. A systematic way to construct candidates is to couple several minimal models through relevant interactions and flow to an IR fixed point. A simple example is three critical 3-state Potts models coupled through their energy operators, whose central charge lies slightly below that of the decoupled theory~\cite{Dotsenko:1998gyp}. For large minimal-model index $m$, these constructions yield perturbative fixed points argued to describe compact unitary irrational CFTs with $c>1$ and only Virasoro chiral symmetry~\cite{Antunes:2022vtb,Antunes:2024mfb,Antunes:2025erb} (see also~\cite{Antunes:2026jjf} for anisotropic Lifshitz fixed points obtained in the same way). Candidate irrational CFTs have also been realised in lattice models with non-invertible symmetries~\cite{Antunes:2025huk}, and the phase transitions of coupled Ising models have been studied in spin chains~\cite{Fuji:2026dvs}. Studying such theories non-perturbatively requires a bootstrap that does not rely on rationality.

This became possible with the numerical conformal bootstrap, introduced in~\cite{Rattazzi:2008pe} and soon applied in $d=2$~\cite{Rychkov:2009ij}, where the bound on the leading scalar dimension displays a kink at the Ising model (see~\cite{Poland:2018epd} for a review). These studies used only the global conformal blocks, valid in any $d$, and therefore did not exploit the full Virasoro symmetry. In $2d$, the latter has been incorporated mostly through the modular bootstrap of torus partition functions~\cite{Hellerman:2009bu,Friedan:2013cba,Collier:2016cls,Fitzpatrick:2023lvh,Benjamin:2026lbj}, recently extended to candidate non-rational CFTs with non-invertible symmetries~\cite{Lin:2023uvm,Ferragatta:2026ejs}, while four-point bootstrap studies with Virasoro blocks have been comparatively scarce~\cite{Lin:2015wcg,Collier:2017shs,Besken:2021eps}. One reason is that no general closed-form expression for Virasoro blocks in terms of standard special functions is known for generic $c$ and weights. Zamolodchikov derived two recursion relations for them, both of which reconstruct the blocks from their poles and residues. The first treats the block as a function of the central charge and expands it around its large-$c$ limit, the global block~\cite{Zamolodchikov:1984eqp}. The second treats it as a function of the exchanged dimension and yields an expansion in the elliptic nome, with rapid convergence near the crossing-symmetric point~\cite{Zamolodchikov:1987avt}. See~\cite{Perlmutter:2015iya} for a nice summary and~\cite{Cho:2017oxl} for further extensions. Evaluating the blocks with the precision and efficiency required by numerical methods is in any case quite costly~\cite{Chen:2017yze}.

Recent progress in the numerical bootstrap in higher dimensions has changed this situation. Efficient methods have been developed to generate global conformal blocks: recursion relations in the spirit of Zamolodchikov's, which build the blocks from their poles in the exchanged dimension~\cite{Kos:2013tga,Penedones:2015aga}, dedicated software for general spinning blocks~\cite{Erramilli:2020rlr}, and rational approximations obtained by optimal interpolation~\cite{Chang:2025mwt}. These methods can be adapted to Virasoro blocks, making it possible to evaluate them efficiently and with the precision required by numerical methods. 

The purpose of this work is twofold: to give a high-level description of an algorithm and code for Virasoro blocks designed for the numerical bootstrap, and to use it to explore several directions in the space of $2d$ CFTs.

As a first application, which also serves as a check of the algorithm, we bootstrap the four-point function of the spin fields of three decoupled copies of the Ising model, organised as a vector of the hypercubic group. Our bootstrap setup only applies for $c>1$: for $c\leq1$ some of the poles of the Virasoro blocks lie on the positive real axis, at physically allowed dimensions. We therefore cannot study the Ising model directly. The tensor product of three copies, with $c=3/2$, is the simplest theory built from it with $c>1$, and its CFT data are known exactly. Imposing gaps compatible with its spectrum, we find a small isolated island around the decoupled point $(\Delta_\sigma,\Delta_S)=(1/8,1)$, where $S=\eps_1+\eps_2+\eps_3$ is the leading scalar singlet.

Next, we continue the programme we initiated in~\cite{Kousvos:2024dlz}, where we studied CFTs with $S_n\ltimes(S_Q)^n$ global symmetry. These are realised by $n$ critical $Q$-state Potts models coupled through their energy operators, $\delta S=g\int d^2x\sum_{a\neq b}\eps_a\eps_b$. This deformation flows to a non-trivial IR fixed point for $n>2$, which is under perturbative control for $Q=2+\epsilon$ and has been studied with transfer matrices and Monte Carlo simulations for $n=Q=3$~\cite{Dotsenko:1998gyp}. For $n=Q=3$, its central charge, $c\approx2.38$, lies slightly below that of the decoupled theory, $c=3 \times 4/5 = 2.4$. The resulting theory is expected to be unitary, compact, irrational, non-supersymmetric and non-multicritical, and is thus among the simplest candidates for an irrational CFT, with no known exact solution. Using global conformal blocks, in~\cite{Kousvos:2024dlz} we found an allowed region for the lowest scaling dimensions that contains the lattice predictions. With Virasoro blocks, our goal is either to gather evidence for the existence of one of the simplest irrational CFTs or to rule it out altogether.

Finally, we move to more exploratory territory. An old conjecture states that there are no full-fledged CFTs in two dimensions with central charge $c=1+\epsilon$, with $\epsilon$ small. How small is part of the question. Unitary CFTs with $c<1$ are minimal models, which accumulate at $c=1$ from below, and the known examples closest to $c=1$ from above, such as the $\mathbb{Z}_5$ parafermion theory at $c=8/7$~\cite{Fateev:1985mm}, lie at a finite distance from it. The conjecture is based on this lack of examples rather than on a proof, so it is worth testing with the bootstrap. Bounding the gap in the $\mathbb{Z}_2$-even sector of the four-point function of a $\mathbb{Z}_2$-odd scalar, we find evidence supporting it: just above $c=1$ the bound collapses to almost zero over a wide range of external dimensions, which excludes any theory whose lightest scalar is absent from its own OPE and lies in this range. Outside it, the bound retains only features of the $c=1$ theories: a narrow spike at $\Delta_\sigma=1/8$, for which we find evidence of a connection to the orbifold branch of the compact boson, and a plateau near $\Delta_\eps=2$, the dimension of its marginal operator. The excluded interval is still present at $c=8/7$, but is absent at $c=7/5$. Further known theories appear between these central charges, above the $\mathbb{Z}_5$ parafermions and their orbifold at $c=8/7$.

\paragraph{Note added.} During the final stages of this project, we became aware that two other groups had independently developed codes to generate Virasoro conformal blocks for the numerical bootstrap~\cite{Poland:2026zyr, Nakayama:2026tba}. We are grateful to Yu Nakayama for informing us of his work and for kindly agreeing to coordinate its submission with ours.

\section{Virasoro conformal blocks}

\subsection{Setup and notation}
We consider a unitary, parity-invariant $2d$ CFT on the plane/sphere, with equal left and right central charges $c_L=c_R=c$. We label positions of local operators by $x = (x^1, x^2) \in \R^2$ or equivalently by complex coordinates $z = x^1 + i x^2$, $\bar{z} = x^1 - i x^2$.
A Virasoro primary of weights $(h,\bar h)$ generates a representation of the left and right Virasoro algebras, with scaling dimension $\Delta=h+\bar h$ and signed $SO(2)$ spin $h-\bar h$. We use $\ell\in\mathbb{Z}_{\geq0}$ for the non-negative spin label. For $\ell>0$, we group the parity-related representations of weights $(h_+,h_-)$ and $(h_-,h_+)$, where $h_\pm=(\Delta\pm\ell)/2$, into a single spin-$\ell$ multiplet. For $\ell=0$, there is only one representation, with weights $(\Delta/2,\Delta/2)$.

A four-point function of parity-even scalar Virasoro primaries $\phi_i(x)$ of dimension $\Delta_i$, with $h_i=\bar h_i=\Delta_i/2$, admits a conformal block decomposition in terms of Virasoro blocks $\cG_{\Delta,\ell}$ in the $(12)(34)$ OPE channel:
\begin{equation}\label{eq:four-point}
    \braket{\phi_{1}(x_1) \phi_2(x_2) \phi_3(x_3) \phi_4(x_4)} = \frac{\big(\tfrac{x_{24}}{x_{14}}\big)^{\Delta_{12}^{-}}\big(\tfrac{x_{14}}{x_{13}}\big)^{\Delta_{34}^{-}}}{x_{12}^{\Delta_{12}^{+}}x_{34}^{\Delta_{34}^{+}}} \sum_{\cO} \lambda_{12\cO} \lambda_{34\cO} \cG_{\Delta_\cO,\ell_\cO}^{12;34}(\z,\zb) \ec
\end{equation}
Here $\Delta_{ij}^{\pm} = \Delta_i \pm \Delta_j$ and $x_{ij} = |x_i - x_j|$. After extracting the displayed kinematical prefactor, the remaining function depends only on the cross ratios
\begin{equation}
    u = \z \, \zb = \frac{x_{12}^{2}x_{34}^{2}}{x_{13}^{2}x_{24}^{2}} \ec \qquad
    v = (1-\z)(1-\zb) = \frac{x_{14}^{2} x_{23}^{2}}{x_{13}^{2} x_{24}^{2}} \ed
\end{equation}
The OPE coefficients $\lambda_{ij \cO}$ are the three-point coefficients for $\braket{\phi_i \phi_j \cO}$, normalised as in~\cite{Poland:2018epd}.\footnote{
    In two-dimensional notation, three-point coefficients $c_{ijk}$ are defined for primaries $O_i(z_i,\bar z_i)$ of conformal weights $(h_i,\bar h_i)$ by
    \[
        \braket{O_1 O_2 O_3} = c_{123} z_{12}^{-h_{123}} z_{23}^{-h_{231}} z_{13}^{-h_{132}}\bar{z}_{12}^{-\bar{h}_{123}} \bar{z}_{23}^{-\bar{h}_{231}} \bar{z}_{13}^{-\bar{h}_{132}} \ec
    \]
    where $h_{ijk} = h_i + h_j - h_k$, and similarly for $\bar h_{ijk}$. For two scalars and a spin-$\ell$ primary, choosing null polarisations $w^\pm=(i,\pm1)/\sqrt2$ gives
    \[
        c_{ij \cO}^{\pm} =  \left(-i/\sqrt{2}\right)^{\ell} \lambda_{ij \cO} \ec
    \]
    where $\pm$ labels the primary of weights $((\Delta\pm\ell)/2,(\Delta\mp\ell)/2)$ of the third operator.
}
The sum runs over exchanged Virasoro primaries appearing in both $\phi_1\times\phi_2$ and $\phi_3\times\phi_4$, counting each parity pair once.
The blocks depend on all four external dimensions, indicated by the superscript $12;34$, and on the central charge $c$, whose dependence we suppress.
We normalise the blocks in the OPE limit as
\begin{equation}
    \cG_{\Delta,\ell}^{12;34}(\z,\zb) \sim \frac{1}{(-2)^{\ell}(1+\delta_{\ell,0})} \left(\z^{\frac{\Delta+\ell}{2}} \zb^{\frac{\Delta-\ell}{2}} + \z^{\frac{\Delta-\ell}{2}} \zb^{\frac{\Delta+\ell}{2}}\right) \ec \qquad z,\bar{z} \to 0 \ec
\end{equation}
matching the higher-dimensional conventions of~\cite{Poland:2018epd}.
Virasoro conformal blocks admit a factorisation
\begin{equation}\label{eq:virasoro-block-full}
    \cG_{\Delta,\ell}^{12;34}(\z,\zb) = \frac{1}{(-2)^{\ell}(1+\delta_{\ell,0})} \left(V_{\frac{\Delta+\ell}{2}}^{12;34}(\z) V_{\frac{\Delta-\ell}{2}}^{12;34}(\zb) + V_{\frac{\Delta-\ell}{2}}^{12;34}(\z) V_{\frac{\Delta+\ell}{2}}^{12;34}(\zb)\right) \ec
\end{equation}
in terms of the chiral blocks $V_h^{12;34}(z)$, which are then normalised as $V_h^{12;34}(z)\sim z^h$ in the OPE limit.

The chiral Virasoro blocks do not admit closed-form expressions, apart from special cases. Nevertheless, they can be computed using Zamolodchikov's elliptic recursion~\cite{Zamolodchikov:1987avt}. We factor out the known large-$h$ asymptotic form, defining a function $H$ that approaches $1$ as $h\to\infty$ at fixed external weights, $c$, and $q$:
\begin{equation}\label{eq:chiral-prefactor}
    V_{h}^{12;34}(\z) = (16 q)^{h-\frac{c-1}{24}}\z^{\frac{c-1}{24}}(1-\z)^{\frac{c-1}{24}-h_{2}-h_{3}} \vartheta_{3}(q)^{\frac{c-1}{2}-4(h_{1}+h_{2}+h_{3}+h_{4})} H(\{h_{i}\}, h, c; q) \ec
\end{equation}
where \( q \) is the elliptic nome
\begin{equation}\label{eq:elliptic-nome}
    q(\z) = \exp\left(-\pi\frac{K(1-\z)}{K(\z)}\right) \ec
\end{equation}
with \( K(\z) = \frac{\pi}{2} \hypgeo{2}{1}\left(\frac{1}{2}, \frac{1}{2}, 1; \z\right) \) the complete elliptic integral of the first kind, and \( \vartheta_{3}(q) = \sum_{n\in \mathbb{Z}} q^{n^{2}} \) the Jacobi theta function.
The function \( H \) then satisfies the recursion relation
\begin{equation}\label{eq:H-recursion}
    H(\{h_{i}\}, h, c; q) = 1 + \sum_{m,n=1}^{+\infty} \frac{R_{m,n}(\{h_{i}\}, c)}{h - h_{m,n}(c)} q^{mn}H(\{h_{i}\}, h_{m,n}(c) + mn, c; q) \ed
\end{equation}
The poles of the recursion are given by
\begin{equation}
    h_{m,n}(c) = - \frac{b^{2}}{4}(m^{2}-1)-\frac{1}{4b^{2}}(n^{2} - 1) - \frac{1}{2}(mn-1) \ec
\end{equation}
and \( b = b(c) \) is a convenient parametrisation of the central charge
\begin{equation}\label{eq:central-charge-parametrisation}
    c = 13 + 6 \left(b^{2} + \frac{1}{b^{2}}\right) \ed
\end{equation}
The residues are given by
\begin{equation}\label{eq:zamolodchikov-residues}
    R_{m,n}(\{h_{i}\}, c) =
    2 \frac{%
        \prod_{p,q}
        \left(\lambda_2 + \lambda_1 - \lambda_{p,q}\right)
        \left(\lambda_2 - \lambda_1 - \lambda_{p,q}\right)
        \left(\lambda_3 + \lambda_4 - \lambda_{p,q}\right)
        \left(\lambda_3 - \lambda_4 - \lambda_{p,q}\right)
    }
    {%
        \prod_{k,l}' \lambda_{k,l}
    } \ec
\end{equation}
where the product in the numerator runs over \( p = -m+1,-m+3,\ldots,m-1 \) and \( q = -n+1,-n+3,\ldots,n-1 \), while the product in the denominator runs over \( k = -m+1,-m+2,\ldots,m \) and \( l = -n+1,-n+2,\ldots,n \) with the pairs \( (k,l) = (0,0), (m,n) \) excluded. The parameters are
\begin{equation}\label{eq:lambda-parameters}
    \lambda_{p,q} = \frac{1}{2}\left(b\,p + \frac{q}{b}\right) \ec \qquad
    \lambda_{i} = \sqrt{\lambda_{1,1}^{2} - h_{i}} \ed
\end{equation}

The parametrisation~\eqref{eq:central-charge-parametrisation} has a branch ambiguity. We pick the solution
\begin{equation}\label{eq:b-branch}
    b(c) = \sqrt{\frac{c-1}{24}} + \sqrt{\frac{c-25}{24}} \ed
\end{equation}
The reciprocal branch is \( 1/b \). The remaining two solutions differ by an overall sign of \( b \), which leaves the recursion unchanged. For \( c > 25 \) the two reciprocal branches are real, for \( 1 < c < 25 \) they are complex of unit modulus and hence complex conjugate, while for \( c < 1 \) they are purely imaginary. The recursion formula is symmetric under \( b \to 1/b \) and \( (m,n) \to (n,m) \) so that \( H \) is a well-defined function of \( c \) and the branch choice is conventional.\footnote{This is manifest in the recursive representation with poles in $c$~\cite{Zamolodchikov:1984eqp}.}

With either \( h_{1} = h_{2} \) or \( h_{3} = h_{4} \), the residues with odd \( mn \) vanish. This follows because the numerator product then contains the factor \( -\lambda_{0,0}=0 \). Thus only positive integer pairs \( m,n\geq1 \) with even \( mn \) need to be retained.

The recursion relation~\eqref{eq:H-recursion} can be used to compute the coefficients of a power series representation of \( H \) in the \( q \) variable
\begin{equation}\label{eq:H-power-series}
    H(\{h_{i}\},h,c;q) = \sum_{N=0}^{\infty} A_{N}(\{h_{i}\}, h, c) q^{N} \ed
\end{equation}
The coefficients are rational functions of $h$ and take the form
\begin{equation}
    \begin{aligned}
      A_{0}(\{h_{i}\},h,c) &= 1 \ec \\
      A_{N}(\{h_{i}\}, h, c) &= \sum_{m n \leq N} \frac{R_{m,n}(\{h_{i}\},c)}{h - h_{m,n}(c)}C_{m,n}^{N}(\{h_{i}\}, c) \ec \qquad (N \geq 1) \ed
    \end{aligned}
\end{equation}
The auxiliary coefficients are shifted-block coefficients,
\begin{equation}\label{eq:C-definition}
    C_{m,n}^{N}(\{h_i\},c)
    = A_{N-mn}(\{h_i\},h_{m,n}(c)+mn,c) \ec \qquad N\geq mn \ec
\end{equation}
and we set \( C_{m,n}^{N}=0 \) for \( N<mn \). Substituting the expression for \( A_{N-mn} \) gives
\begin{equation}\label{eq:C-recursion}
    \begin{aligned}
      C_{m,n}^{mn}(\{h_{i}\},c) &= 1 \ec \\
      C_{m,n}^{N}(\{h_{i}\}, c) &= \sum_{rs \leq N - mn} \frac{R_{r,s}(\{h_{i}\},c)}{h_{m,n}(c) + mn - h_{r,s}(c)} C_{r,s}^{N-mn}(\{h_{i}\},c) \ec \qquad (mn < N)\ed
    \end{aligned}
\end{equation}
The shift in the last coefficient is \( N-mn \), because the outer recursion term has already supplied the \( q^{mn} \) power. All sums over pole labels have positive integer indices. Following~\cite{Chen:2017yze}, we can implement the computation of the shifted-block coefficients efficiently in increasing power-series order, grouping poles with the same product \( mn \) so that each update is a matrix--vector multiplication.
These expressions are understood at generic parameter values. When individual recursion terms become singular, the corresponding limits must be taken after resolving cancellations~\cite{Chen:2017yze}.

\subsection{Rational approximation}
When either external pair has equal weights, odd-$mn$ labels are omitted from all denominator products below, and $K$ counts the remaining labels.
At fixed $z=z_*$, truncating the power series~\eqref{eq:H-power-series} at order $q^N$ gives an approximation to the chiral Virasoro block as an exponential factor times a rational function of $h$. Its derivatives in $z$ have the same denominator and exponential factor, with a numerator polynomial $P_n(h)$ for the $n$-th derivative:
\begin{equation}\label{eq:rational-approx-chiral}
    \partial_z^n V_{h}^{12;34} \approx \frac{(16 q_*)^h}{D(h)} P_n(h) \ec \qquad n\geq0 \ec
\end{equation}
where
\begin{equation}
    D(h) = \prod_{(m,n)}^{mn \leq N}(h - h_{m,n}(c)) \ec
\end{equation}
For a fixed set of retained pole labels, the denominator depends on $c$, while the numerator also depends on the external weights and $z_*$. We suppress these dependencies for simplicity.
For bootstrap applications, we will choose $z_{*} = 1/2$, which means $q_* = e^{-\pi} \approx 0.043$.

Combining~\eqref{eq:rational-approx-chiral} with~\eqref{eq:virasoro-block-full} gives rational approximations to the Virasoro blocks and their derivatives as functions of $\Delta$:
\begin{equation}\label{eq:rational-approx-full}
    \partial_{\z}^{m}\partial_{\zb}^{n}\mathcal{G}_{\Delta,\ell}^{12;34} \approx \frac{(16 q_{*})^{\Delta}}{\mathcal{D}_{\ell}(\Delta)} \mathcal{P}^{m,n}_{\ell}(\Delta) \ec
\end{equation}
where
\begin{equation}
    \mathcal{D}_{\ell}(\Delta) = \prod_{(r,s)}^{rs \leq N} (\Delta + \ell - 2 h_{r,s}(c))(\Delta - \ell - 2 h_{r,s}(c)) \ec
\end{equation}
and
\begin{equation}\label{eq:full-block-polynomial}
    \mathcal{P}^{m,n}_{\ell}(\Delta) = \frac{2^{2K}}{(-2)^\ell(1+\delta_{\ell,0})}\Big(P_{m}\big(\tfrac{\Delta+\ell}{2}\big)P_{n}\big(\tfrac{\Delta-\ell}{2}\big) + P_{m}\big(\tfrac{\Delta-\ell}{2}\big)P_{n}\big(\tfrac{\Delta+\ell}{2}\big)\Big) \ec
\end{equation}
where $K$ counts the retained positive integer pairs $(r,s)$ with $rs\leq N$. The polynomial degrees satisfy
\[
    \deg P_n\leq K+n \ec \qquad \deg\mathcal P^{m,n}_{\ell}\leq2K+m+n \ed
\]

\subsection{Numerical bootstrap and implementation}

With the rational approximations at hand, we can implement a numerical conformal bootstrap for $2d$ CFTs with Virasoro symmetry. Crossing symmetry of~\eqref{eq:four-point} in the simplest case of identical operators gives the sum rule
\begin{equation}\label{eq:sum-rule}
    \sum_{\cO} \lambda_{\phi \phi \cO}^2 \cF^{\phi\phi;\phi\phi;-}_{\Delta_{\cO},\ell_{\cO}}(z,\bar{z}) = 0 \ec
\end{equation}
where the crossing functions are
\begin{equation}\label{eq:crossing-function}
    \cF^{12;34;\pm}_{\Delta,\ell}(z,\bar{z}) = (\z \zb)^{-\frac{\Delta_{12}^{+}}{2}} \cG^{12;34}_{\Delta,\ell}(z,\bar{z}) \pm ((1-\z) (1-\zb))^{-\frac{\Delta_{12}^{+}}{2}} \cG^{12;34}_{\Delta,\ell}(1-z,1-\bar{z}) \ed
\end{equation}
For mixed correlators, this convention corresponds to crossing under $1\leftrightarrow3$, relating the orderings $12;\! 34$ and $32;\! 14$.

Compared to the numerical bootstrap used in higher dimensions, or in $2d$ with global conformal blocks, the bootstrap problem now depends explicitly on the central charge $c$ of the CFT, which enters as a parameter in the conformal blocks. Moreover, the sum over Virasoro primaries in~\eqref{eq:sum-rule} does not contain the stress tensor, since its contribution is automatically taken into account in the Virasoro block of the identity.

Derivatives of the prefactor-dressed blocks entering~\eqref{eq:crossing-function} follow by the Leibniz rule, with the same denominator and exponential factor as in~\eqref{eq:rational-approx-full}.
We use linear functionals built from derivatives with $m+n\leq\Lambda$ at the crossing-symmetric point, imposing the standard bootstrap positivity conditions to bound the spectrum and OPE coefficients (see e.g.~\cite{Chester:2019lectures} for lecture notes). For mixed correlators, OPE coefficients enter quadratic forms, and the functional constraints require positive semi-definite matrices~\cite{Kos:2014bka}.
The rational approximation converts these constraints into a polynomial matrix program (PMP), solved using \texttt{SDPB}~\cite{SimmonsDuffin:2015SDPB,Landry:2019SDPB}, by removing the positive prefactor. For $c>1$ and identical external operators, the denominator is positive throughout the unitary region
\begin{equation}
    \mathcal{D}_{\ell}(\Delta) > 0 \ec \qquad \Delta \geq \ell \ed
\end{equation}
Indeed, after omitting the vanishing odd-$rs$ residues in the identical-operator setup, the poles $h_{r,s}(c)$ are strictly negative for $c\geq25$. For $1<c<25$, the diagonal poles remain negative, while off-diagonal poles occur in complex-conjugate pairs. For generic mixed correlators, the pole $h_{1,1}=0$ can remain, so strict positivity holds for $\Delta>\ell$. For $0<c\leq1$, poles can lie in the unitary region $\Delta \geq \ell$, producing zeros of the denominator and obstructing this formulation.
Our numerical bootstrap setup will apply only for $c > 1$.\footnote{
    For $c < 1$, unitary CFTs are in any case already classified by minimal models.
}

A technical point, but important in practice, is that the degree of the polynomials in the rational approximation~\eqref{eq:rational-approx-full} grows as $O(N\log N)$ at fixed derivative order.\footnote{
    For example, at $N=50$ and total derivative order up to $20$, the numerator degree is at most $434$, or $298$ when either external pair has equal weights. The recursion order required to achieve a target relative truncation error depends on the central charge, external and exchanged dimensions, and derivative order.
}
This makes a naive PMP implementation impractical.
This problem is well known for the rational approximation to conformal blocks in higher dimensions (although less severe), and was first addressed by ``pole shifting''~\cite{Kos:2013tga}, matching derivatives at the unitarity bound and asymptotic data at large $\Delta$, and more recently by polynomial interpolation~\cite{Chang:2025mwt}.
Polynomial interpolation allows for better control of the approximation over the full positivity domain using a weighted norm.
We implement the polynomial interpolation in our setup, which makes the numerical bootstrap manageable on a laptop (at small derivative order) or on a cluster.
We choose a retained-pole cutoff $\kappa\leq N$, keeping the labels with $rs\leq\kappa$ in each chiral factor. If $K_\kappa$ is their number, counted with multiplicity, we use a polynomial of degree at most $2K_\kappa+\Lambda$ and $2K_\kappa+\Lambda+1$ interpolation nodes. This choice matches the maximal large-$\Delta$ power of the rational target. We perform interpolation with respect to the weighted norm
\begin{equation}
    \norm{f} = \sup_{x \geq 0} \mu_{\kappa}(x) \abs{f(x)} \ec \qquad \mu_{\kappa}(x) = \frac{r^x}{\prod_{i}(x + k_i)} \ec
\end{equation}
where $x=\Delta-\Delta\ped{min}$ measures the dimension above the assumed gap in a positivity condition, $r=16q_*$,\footnote{
    Different exponential weights can change the interpolation error. The large-$\Delta$ asymptotics of the Virasoro OPE coefficients suggest that using $r=q_*$ in the error norm might improve the approximation to the action of optimal bootstrap functionals~\cite{Chang:2025mwt}. We have not addressed this matter systematically.
} and the parameters $k_i$ run over $\Delta\ped{min}\pm\ell-2h_{r,s}(c)$ with $rs\leq\kappa$, including both signs and all multiplicities. Labels with vanishing odd-$rs$ residues are omitted when either external pair has equal weights. The poles of the weight are at $x=-k_i$.
Here we assume that no retained pole lies on the allowed domain. Writing $\mu_N$ for the same weight with cutoff $N$, we interpolate the function
\begin{equation}
    \frac{\mu_{N}(x)}{\mu_{\kappa}(x)} \cP^{m,n}_{\ell}(\Delta\ped{min}+x) \ec
\end{equation}
by a polynomial $\widetilde{\mathcal P}^{m,n}_\ell(x)$, using the interpolation nodes in~\cite[Eqs.\ 3.48--3.49]{Chang:2025mwt}.\footnote{
    Complex-conjugate pairing for $1 < c < 25$ makes the weight real and positive on the allowed domain. The construction in~\cite{Chang:2025mwt} assumes a density supported on a single interval $[0,b]$, with $b>0$, and yields real, ordered nodes when this density is non-negative and satisfies the equilibrium conditions.
    This is consistent for global conformal blocks and for Virasoro blocks with $c\geq13$ and $\Delta\ped{min}\geq\ell$, since the poles have negative real parts.
    For $1<c<13$, some poles can have positive real parts, and the candidate density may be negative or fail to satisfy the equilibrium conditions. 
    The equilibrium density still exists, but its support may be an interval $[a,b]$ with $a>0$ or several intervals, and can be found numerically.
    In the parameter sets tested, violations of the single-interval equilibrium conditions were found only close to $c=1$, without establishing a universal threshold. 
    Small interpolation errors were also obtained in tested configurations where these conditions failed.
}
The resulting block approximation is
\begin{equation}\label{eq:interpolated-block}
    \partial_z^m\partial_{\bar z}^n\mathcal G_{\Delta,\ell}^{12;34}
    \approx r^{\Delta\ped{min}}\mu_\kappa(x)\widetilde{\mathcal P}^{m,n}_\ell(x) \ed
\end{equation}
The weighted interpolation error therefore measures the additional absolute error in the rational block approximation, up to the constant $r^{\Delta\ped{min}}$. The same construction applies to derivatives of the prefactor-dressed blocks entering the crossing functions.

The code was tested against several examples of known Virasoro blocks, which we collect in Appendix~\ref{app:known-blocks} for reference. The comparison has to be done with some care since the known Virasoro blocks involve degenerate operators where the recursion is singular (see~\cite{Javerzat:2018maf} for a detailed discussion). We approach the degenerate parameters along a regulated path and verify convergence to the exact block as the regulator decreases, controlling recursion truncation and numerical precision separately. We also checked our code against the low-order power-series coefficients reported in analytic form in~\cite{Mironov:2010zs}. The implementation was also checked against the code developed for~\cite{Chen:2017yze}.
Recursion truncation, interpolation, and finite-precision evaluation introduce separate errors. Their control requires varying $N$, $\kappa$, and the working precision independently.

The implementation of the rational approximation and polynomial interpolation was carried out in \texttt{Mathematica} and integrated into the \texttt{simpleboot} software~\cite{simpleboot}.
It is made publicly available at the repository \url{https://gitlab.com/stek12/virasoro_bootstrap}.
The repository contains:
\begin{itemize}
    \item \texttt{Packages/VirasoroBlocks.m}: library file implementing the recursion relation and rational approximation of derivatives.
    \item \texttt{Packages/VirasoroBlocksCli.m}: a command line interface for generating the Virasoro blocks. The script needs as input parameters the central charge $c$, the scaling dimensions of the external operators $\Delta_i$, the order of the recursion relation $r_{N}$,\footnote{Note that internally the order is set to be $N = r_{N} + \Lambda$, as higher derivatives require higher order of the recursion to achieve the same truncation error.} the maximum derivative order $\Lambda$, a list of exchanged spins $\ell$, and the working precision in decimal digits. The script will output rational approximations of the derivatives of $(\z \zb)^{-\Delta_{12}^{+}/2} \cG^{12;34}_{\Delta,\ell}(z,\bar{z})$ at $z=\bar{z}=1/2$ in the same \texttt{Mathematica} format as \texttt{scalar\_blocks}\footnote{\url{https://gitlab.com/bootstrapcollaboration/scalar_blocks}}  and \texttt{scalar\_blocks\_mod}\footnote{\url{https://gitlab.com/suning-git/scalar-blocks-mod}}. A flag \texttt{--bare} can be specified to remove the $(\z \zb)^{-\Delta_{12}^{+}/2}$ prefactor.
    \item \texttt{VirasoroSDPB.m}: a small extension of the polynomial interpolation code in \url{https://github.com/davidsd/sdpb/blob/master/mathematica/SDPB.m} to support the complex poles in certain ranges of central charge.\footnote{
        We warn the user that the input generated for \texttt{pmp2sdp} will contain a ``fake'' list of poles obtained by taking $-|\mathrm{Re}(\ldots)|$ of the poles, because \texttt{pmp2sdp} does not support complex numbers. This does not affect the interpolation since the interpolation nodes are provided directly as part of the input, and the list of poles is discarded.
    }
    \item \texttt{Packages/VirasoroSimpleboot.m}: provides the \texttt{simpleboot} integration for constructing the PMP from the crossing equations, which \texttt{pmp2sdp} parses and \texttt{SDPB} solves.
    \item \texttt{Tutorial/Z2SingleCorrelator.nb}: an example \texttt{simpleboot} notebook for running a gap-maximisation bound, as presented in Section~\ref{sec:cft-desert} (can run on a laptop).
    \item \texttt{Workspace}: collection of various script files used for the bounds obtained in Sections~\ref{sec:ising}, \ref{sec:potts} and~\ref{sec:cft-desert} (meant to be run on a cluster).
\end{itemize}
We warn the reader that the repository is under development and might be subject to changes in the future.

\section{Application 1: Ising island in $2d$}\label{sec:ising}
With the machinery in place, we proceed to the first application, which will also provide a nice cross-check of our setup. Our first target is the Ising CFT. However, since its central charge $c=\frac{1}{2}$ lies below $c=1$, it is not directly accessible in our bootstrap setup. In order to access it, we will consider three decoupled replicas of the theory, which have central charge $c=3/2$ and global symmetry $S_3 \ltimes (\Z_2)^3$. This is the familiar cubic symmetry, long studied in the conformal bootstrap (see e.g.~\cite{Rong:2017cow,Stergiou:2018gjj}). The spectrum in the theory of three decoupled replicas is completely fixed by that of a single Ising CFT and can be computed systematically via a Virasoro character expansion (see Appendix~\ref{app:spectrum}). The leading scalar singlet $S$ can be written as $S=\eps_1+\eps_2+\eps_3$, where $\eps_i$ are the energy operators of each Ising replica. The three spin fields $\sigma_i$, each odd under the corresponding replica's $\Z_2$, form the fundamental representation of $S_3 \ltimes (\Z_2)^3$.

\begin{figure}[t]
  \centering
  \includegraphics[width=0.62\textwidth]{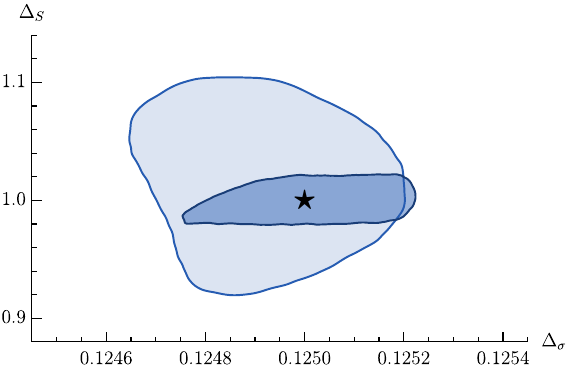}
  \caption{Allowed region for three decoupled Ising models at $c=3/2$, from the four-point function $\langle \sigma_i \sigma_j \sigma_k \sigma_l \rangle $. $\Delta_S$ is the
  dimension of an isolated scalar in the singlet channel ($\eps_1+\eps_2+\eps_3$). The other gaps are set just below the spectrum of three
  decoupled Ising models and lowered by a margin $\delta$ (Table~\ref{tab:ising-assumptions}). Dark: $\Lambda=11$,
  $\delta=0.01$; light: $\Lambda=19$, $\delta=0.1$; the numerical parameters can be found in Table~\ref{tab:ising-parameters}. The star marks the theory of three decoupled Ising models, $(\Delta_\sigma,\Delta_S)=(1/8,1)$.}
  \label{fig:ising-island}
\end{figure}

We then consider the crossing equation for the correlator $\langle \sigma_i \sigma_j \sigma_k \sigma_l \rangle$, already treated for $d=3$ in~\cite{Stergiou:2018gjj}. The $\sigma_i\times\sigma_j$ OPE decomposes into the singlet $S$, traceless diagonal $X$, off-diagonal symmetric $Z$, and antisymmetric $B$ representations; the first three contain even spins, while $B$ contains odd spins. In order to try and isolate the theory, we impose gap assumptions dictated by the spectrum of the decoupled theory, within a margin $\delta$ when stronger than unitarity (see Table~\ref{tab:ising-assumptions}). In particular, other than for the first scalar singlet, we do not demand that any other operator in the spectrum is isolated. We show the constraints obtained for $\Lambda=11$ and $\Lambda=19$ in Figure~\ref{fig:ising-island}. Somewhat counterintuitively, the smaller island is for $\Lambda=11$ and the bigger one for $\Lambda=19$. This is because in the former case we use $\delta=0.01$ (stricter gaps) and in the latter $\delta=0.1$ (less constraining gaps). The reason for this choice is that the $\Lambda=19$ plot exhibited the phenomenon known as stalling in the SDPB runs; in other words, we did not see primal or dual jumps. A related difficulty of the two-dimensional crossing equations at finite derivative order was first noted in the context of central-charge bounds~\cite{Vichi:2011zza}; see also the discussion in~\cite{Chester:2019ifh}. We attribute this to truncation artifacts given the rather aggressive nature of the gaps we have chosen and do not interpret it as a physics effect.

\begin{table}[t]
  \centering\small
  \begin{tabular}{c c l}
    \toprule
    channel & spin & assumption \\ \midrule
    $S$ & 0 & $\Delta = \Delta_S$ and $\Delta\geq4-\delta$ \\
    $S$ & 2 & $\Delta\geq6-\delta$ \\
    $S$ & $4,6,\dots$ & $\Delta\geq\ell$ \\
    $X$ & 0 & $\Delta\geq1-\delta$ \\
    $X$ & $2,4,\dots$ & $\Delta\geq\ell$ \\
    $Z$ & $0,2,\dots$ & $\Delta\geq\ell+\tfrac14-\delta$ \\
    $B$ & $1,3,\dots$ & $\Delta\geq\ell+\tfrac14-\delta$ \\
    \bottomrule
  \end{tabular}
  \caption{Spectrum assumptions in the channels of $\sigma_i\times\sigma_j$ for Figure~\ref{fig:ising-island}. The
  dimensions $\Delta_\sigma$ and $\Delta_S$ are scanned over. We require the lightest singlet scalar to be isolated, with a gap above it. Each gap is the lowest operator of three decoupled Ising
  models in that channel and spin (Table~\ref{tab:ising-spectrum}), lowered by a margin $\delta$.}
  \label{tab:ising-assumptions}
\end{table}

\section{Application 2: coupled vs decoupled Potts models}\label{sec:potts}

\begin{figure}[t]
  \centering
  \includegraphics[width=0.62\textwidth]{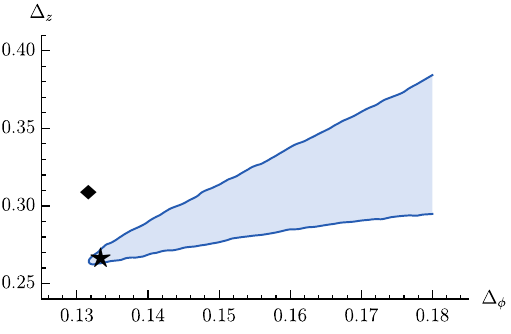}
  \caption{Single correlator bound $\braket{\phi\phi\phi\phi}$ (6 crossing equations, in the channels $S^+$, $X^+$, $V^+$,
  $Z^+$, $A^-$, $B^-$) at $\Lambda=11$ and $c = 2.4$. $\phi$ and $z$ are isolated scalars in the $V^+$ and $Z^+$ channels. All channels
  satisfy the gaps of Table~\ref{tab:potts-gaps-decoupled}, lowered by $\delta=0.01$.
  The allowed region is a thin peninsula whose tip is located next to the theory of three decoupled Potts models (star); the coupled theory (diamond) is
  excluded.}
  \label{fig:potts-single}
\end{figure}

Having seen that the method can produce bootstrap islands for $c > 1$ in the previous section, we now study the replica-Potts theory with $S_3 \ltimes(S_3)^3$ global symmetry, originally treated with the bootstrap and global conformal blocks in \cite{Kousvos:2024dlz}.

The sum rules for the mixed $\phi$--$z$ crossing system were derived by constructing the
Clebsch--Gordan coefficients of $S_3 \ltimes (S_3)^3$ explicitly, by averaging over the group in a fixed
basis of each irreducible representation, building the four-point tensor structures from them, and
imposing crossing in a basis of four-point invariants (this implementation differs from~\cite{Go:2019lke} in the way the Clebsch--Gordan coefficients are calculated). Then, the resulting 57 sum rules were
verified against (generalised) free-field solutions.
The necessary group data can also be found in \textbf{GAP}~\cite{GAP4}.\footnote{\href{https://www.gap-system.org/}{https://www.gap-system.org/}} This algorithm is efficiently implementable via the reader's favourite choice of LLM.

We first study the four-point function of the fundamental field ${\phi}^a_i$ and impose gap assumptions analogous to those used for the decoupled Ising models in the previous section. In particular, we again impose gaps compatible with the spectrum extracted via the character expansion in the decoupled replica limit. This leads to Figure~\ref{fig:potts-single}, where we observe a very sharp peninsula in the allowed region for the scaling dimensions of the leading scalars $\phi$ and $z$ in the $V$ and $Z$ representations.\footnote{In perturbation theory, $z^{ab}_{ij}\sim\phi^a_i \phi^b_j$.} The sharp tip lies near the decoupled theory point. Notice that the coupled theory point, as determined via transfer matrices in~\cite{Dotsenko:1998gyp}, is excluded. This is expected, since various operators gain (significant) negative anomalous dimensions in the interacting theory. See, for example, the summary of known data in Table 3.1 of~\cite{Kousvos:2024dlz}. In particular, the leading $X$ operator (in the perturbative language: $\eps_i -\frac{1}{3}(\eps_1 +\eps_2+\eps_3)$) has a dimension of $\Delta_X=0.63 \pm 0.03$ from transfer matrices~\cite{Dotsenko:2001cct} and $\Delta_X=0.63 \pm 0.04$ from Monte Carlo~\cite{Dotsenko:1998gyp}. From this plot, we learn that while the bootstrap is still particularly constraining, it is no longer strong enough (at this derivative order) to provide an island as in the Ising case.

\begin{figure}[t]
  \centering
  \includegraphics[width=0.62\textwidth]{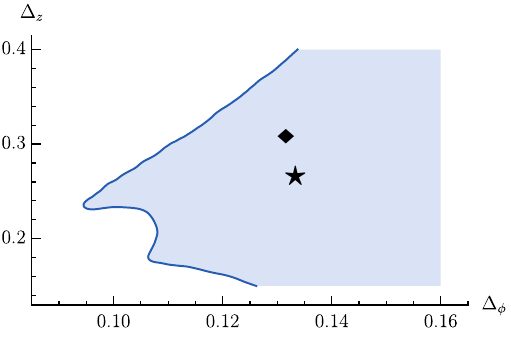}
  \caption{Allowed region in the $\Delta_\phi$--$\Delta_z$ plane, from the mixed $\phi$ and $z$ crossing equation system (57 sum rules) at $\Lambda=11$ and $c = 2.4$. The gaps are those of
  Table~\ref{tab:potts-gaps-mild}; the numerical parameters are given in Table~\ref{tab:potts-parameters}. The star marks the theory of three decoupled Potts models,
  $(\Delta_\phi,\Delta_z)=(2/15,4/15)$, and the diamond the coupled fixed point as estimated via transfer
  matrices~\cite{Dotsenko:1998gyp}. Both lie in the allowed region.}
  \label{fig:potts-generic}
\end{figure}

\begin{table}[t]
  \centering\small
  \begin{tabular}{c c l}
    \toprule
    channel & spin & assumption \\ \midrule
    $S^+$ & 0 & $\Delta\geq\tfrac12$ \\
    $V^+$ & 0 & $\phi$ isolated; all other scalars $\Delta\geq\tfrac12$ \\
    $Z^+$ & 0 & $z$ isolated; all other scalars $\Delta\geq\tfrac45$ \\
    $Z_3^+$ & 0 & $\Delta\geq\tfrac1{10}$ \\
    $X^+$ & 0 & $\Delta\geq\tfrac25$ \\
    \multicolumn{2}{l}{all other channels} & $\Delta\geq\ell+10^{-6}$ \\
    \bottomrule
  \end{tabular}
  \caption{Gap assumptions for Figure~\ref{fig:potts-generic}. Channels are labelled by the irrep of
  $S_3\ltimes(S_3)^3$ (Appendix~\ref{app:spectrum}) and by the spin ($+$: even spins, $-$: odd spins).}
  \label{tab:potts-gaps-mild}
\end{table}

\begin{figure}[t]
  \centering
  \includegraphics[width=0.62\textwidth]{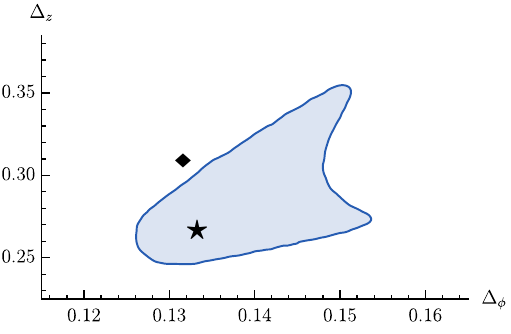}
  \caption{Mixed correlator $\phi$--$z$ island enclosing the decoupled Potts theory and excluding the coupled fixed point. The gaps used are those of Table~\ref{tab:potts-gaps-decoupled}, set by the
  spectrum of three decoupled Potts models (Appendix~\ref{app:spectrum}) and lowered by a margin $\delta=0.1$. The decoupled theory is denoted by a star, and the coupled theory by a diamond.}
  \label{fig:potts-decoupled}
\end{figure}

We thus proceed to the bounds we can obtain from the mixed $\phi$--$z$ correlator system. This consists of a bootstrap system with 57 sum rules.\footnote{The mixed system includes $\braket{\phi\phi\phi\phi}$, $\braket{zzzz}$, $\braket{\phi\phi zz}$, $\braket{\phi z\phi z}$, $\braket{\phi\phi\phi z}$, and $\braket{zzz\phi}$.} While particularly costly to run numerically, it provides another good cross-check of our setup. We start by probing the scaling dimensions in the $\Delta_\phi$--$\Delta_z$ plane with some (very) conservative assumptions (Table~\ref{tab:potts-gaps-mild}), and see that while the bootstrap can carve out an allowed space (Figure~\ref{fig:potts-generic}), it is not strong enough to isolate or exclude either the decoupled or the coupled theory. Tightening the gap assumptions to those of Table~\ref{tab:potts-gaps-decoupled}, in Figure~\ref{fig:potts-decoupled}, we see that the allowed region now collapses to an isolated island which excludes the coupled theory for the reasons explained above (operators gain large negative anomalous dimensions and violate gap assumptions).

This leaves us with the task of locating the coupled theory and ideally obtaining an island for it, which would correspond to the first bootstrap island for a (potentially) non-rational CFT in $2d$. To explore this, firstly note that if all operators in the coupled fixed point had positive anomalous dimensions, we could impose the same gaps that lead to the decoupled island, plus some additional ones motivated by perturbation theory and/or microscopic simulations that single out the interacting theory. We found that the anomalous dimensions of operators in some specific representations are negative (within leading-order conformal perturbation theory), which agrees with the known values for operators like $X$ from independent non-perturbative studies. Thus, relaxing these gaps as in Table~\ref{tab:potts-gaps-coupled}, we should in principle be able to recover the coupled theory. However, doing only this still leaves the decoupled theory as an allowed point, which we would like to get rid of. This can be remedied by raising the gap on the leading scalar singlet $S$ to $\Delta_S \geq1$. At large $n$ (the limit of many replicas, $S_n\ltimes {S_3}^n$) this assumption is justified, since $S$ acquires the shadow dimension of its decoupled value, $\Delta_S^{\text{coupled}}=d-\Delta_S^{\text{decoupled}}=1.2$~\cite{Binder:2021vep}. This large-$n$ value is in fact in good agreement with microscopic simulations for $n=3$ (see Table~3.1 of~\cite{Kousvos:2024dlz}).

\begin{figure}[t]
  \centering
  \includegraphics[width=0.62\textwidth]{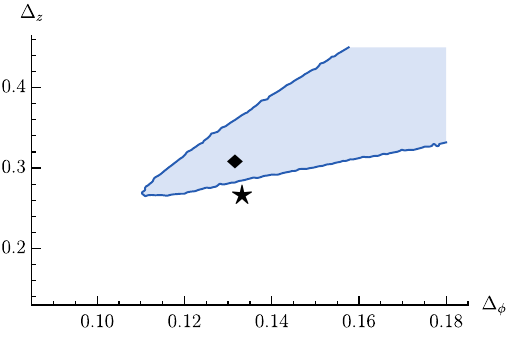}
  \caption{Mixed correlator $\phi$--$z$ peninsula enclosing the coupled Potts theory and excluding the decoupled theory. The gaps used are those of Table~\ref{tab:potts-gaps-decoupled}, supplemented by the modifications of Table~\ref{tab:potts-gaps-coupled}: the gaps
  are relaxed in the channels where the coupled theory is expected to have lighter operators than the decoupled one,
  and the first singlet scalar is required to lie above $\Delta=1$, which removes the leading singlet $\sum_i\eps_i$ of the decoupled
  theory ($\Delta=4/5$). The coupled theory is denoted by a diamond,
  and the decoupled theory by a star.}
  \label{fig:potts-coupled}
\end{figure}

These combined assumptions lead to Figure~\ref{fig:potts-coupled}, which indeed allows for the coupled theory and excludes the decoupled one. While these gap assumptions are to some extent ad hoc, since anomalous dimensions could become negative away from the perturbative limit, we see that the bootstrap has, at least in principle, the possibility to isolate the coupled theory. We also demonstrated that the bootstrap is strong enough to obtain an island for the decoupled theory at $c=2.4$ (which lies far from $c \rightarrow 1$, where, as we will see, Virasoro blocks are most constraining).

The prediction for $c$ in the coupled-Potts theory is around $c=2.377 \pm 0.003$ \cite{Dotsenko:1998gyp}. Our plots therefore do not apply to it directly, although the difference in central charge is small. We believe that these findings, combined with further independent calculations using microscopic simulations or perturbation theory, could provide the spectral information needed to obtain a bootstrap island for the coupled theory.

\begin{table}[t]
  \centering\small
  \begin{tabular}{l c l}
    \toprule
    channels & spins & assumption \\ \midrule
    $S^+$ & 0 & $\Delta\geq1$ \\
    $X^+$ & 0 & $\Delta\geq\tfrac25$ \\
    $B^\pm$, $XV^\pm$, $XA^-$ & all & $\Delta\geq\ell$ \\
    $VB^\pm$, $B_3^-$, $AB^+$ & all & gaps of Table~\ref{tab:potts-gaps-decoupled}, lowered by $\tfrac14$ instead of $\delta$ \\
    \bottomrule
  \end{tabular}
  \caption{Changes with respect to Table~\ref{tab:potts-gaps-decoupled} for Figure~\ref{fig:potts-coupled} in order to single out the coupled theory. All
  other channels keep the gaps of Table~\ref{tab:potts-gaps-decoupled} with $\delta=0.1$.}
  \label{tab:potts-gaps-coupled}
\end{table}

\section{Application 3: a CFT desert for $c\sim 1$}\label{sec:cft-desert}

\subsection{The conjecture}

Unitary CFTs with $c<1$ are completely classified: unitarity restricts the central charge to the discrete series of the minimal models~\cite{Friedan:1983xq}, and every such theory is rational. For compact unitary theories at $c=1$, there is a well-established catalogue, consisting of the compact boson, its orbifold and a few exceptional points~\cite{Ginsparg:1987eb}. For $c>1$, instead, no classification is known. All known compact unitary CFTs are either rational or marginal deformations of rational theories~\cite{Benjamin:2026lbj}, and none of them lies arbitrarily close to $c=1$. The known examples closest to $c=1$ from above are the $\mathbb{Z}_5$ parafermion theory at $c=8/7$~\cite{Fateev:1985mm}, its $\Z_2$ charge-conjugation orbifold~\cite{Jiang:2017orbifold}, and the $m=5$ $\mathcal{N}=1$ superconformal minimal model at $c=81/70$~\cite{Friedan:1984rv}. Liouville theory exists for any $c>1$, but it is non-compact: it has no normalisable $\mathrm{SL}(2,\R)$-invariant vacuum, and its spectrum is continuous, starting at $\Delta=(c-1)/12$~\cite{Ribault:2014hia}. This has led to the conjecture that there is a ``desert'' above $c=1$: no compact unitary CFT exists with $c=1+\epsilon$ for $\epsilon$ small enough, and how small is itself part of the question. As far as we know, the conjecture rests only on the absence of examples, and it is therefore worth testing directly.

The modular bootstrap has addressed this region by constraining the spectral gap $\Delta\ped{gap}$, the dimension of the lightest non-identity Virasoro primary. The universal bound of~\cite{Hellerman:2009bu}, refined in~\cite{Friedan:2013cba,Collier:2016cls}, gives $\Delta\ped{gap}\leq c/6+1/3$ for $1\leq c\leq4$. At $c=1$, this is saturated by the compact boson at the self-dual radius, so on its own it does not exclude theories just above $c=1$. Imposing integer degeneracies improves the bound by a few percent for $c\gtrsim1$~\cite{Fitzpatrick:2023lvh}. More recently,~\cite{Benjamin:2026lbj} attacked the problem from the primal side, numerically constructing truncated modular-invariant partition functions with integer degeneracies. They find candidate spectra throughout $1<c<8/7$, suggesting that modular invariance alone does not forbid CFTs in this range. At the same time, they find no solutions for $1<c<1.06$ with $0.3<\Delta\ped{gap}<0.5$, which they attribute to integrality and which points to a stronger bound on the gap near $c=1$. Modular invariance is a necessary but not sufficient condition, and crossing symmetry of four-point functions provides independent constraints. These are the constraints we study here.

\begin{table}[t]
\centering
\footnotesize
\setlength{\tabcolsep}{3pt}
\begin{tabular}{l l l l l}
\toprule
Theory & $c$ & $\Delta_\sigma$ & $\Delta_\eps$ & Ref. \\
\midrule

$\mathbb{Z}_k$ parafermions, $k\geq5$
& $\frac{2(k-1)}{k+2}$ \ $\big(\frac87\big)$
& $\begin{array}{@{}l@{}}
     \frac{k-1}{k(k+2)} \ \big(\frac4{35}\big)\\[2pt]
     \boldsymbol{\big(\frac12\big)}\\[2pt]
     \frac{4}{k+2} \ \big(\frac47\big)
   \end{array}$
& $\begin{array}{@{}l@{}}
     \frac{2(k-2)}{k(k+2)} \ \big(\frac6{35}\big)\\[2pt]
     \boldsymbol{\big(\frac85\big)}\\[2pt]
     \frac{12}{k+2} \ \big(\frac{12}7\big)
   \end{array}$
& \cite{Fateev:1985mm,Jiang:2017orbifold,Jiang:2019fusion} \\ \addlinespace[5pt]

$\mathcal{N}=1$ MM, $m\geq5$
& $\frac32\big(1-\frac{8}{m(m+2)}\big)$
  \ $\big(\frac{81}{70}\big)$
& $\frac{3}{m(m+2)}$ \ $\big(\frac3{35}\big)$
& $\frac{8}{m(m+2)}$ \ $\big(\frac8{35}\big)$
& \cite{Friedan:1984rv,Fredenhagen:2007tk} \\[3pt]

$\frac{\mathrm{SU}(2)_k\times\mathrm{SU}(2)_l}
       {\mathrm{SU}(2)_{k+l}}$, $k\geq l\geq3$
& $\frac{3k}{k+2}+\frac{3l}{l+2}
   -\frac{3(k+l)}{k+l+2}$ \ $\big(\frac{27}{20}\big)$
& $\frac{3l}{2(k+2)(k+l+2)}$ \ $\big(\frac9{80}\big)$
& $\frac{4l}{(k+2)(k+l+2)}$ \ $\big(\frac3{10}\big)$
& \cite{DiFrancesco:1997nk} \\[3pt]

$\mathrm{SU}(2)_k$ WZW, $k\geq2$
& $\frac{3k}{k+2}$ \ $\big(\frac32\big)$
& $\frac{3}{2(k+2)}$ \ $\big(\frac38\big)$
& $\frac{4}{k+2}$ \ $(1)$
& \cite{Knizhnik:1984nr,DiFrancesco:1997nk} \\[3pt]

$\mathcal{N}=2$ MM, $k\geq2$
& $\frac{3k}{k+2}$ \ $\big(\frac32\big)$
& $\frac{\min(k,4)}{4(k+2)}$ \ $\big(\frac18\big)$
& $\begin{array}{@{}l@{}}
     \frac{k}{k+2}\quad(k=2,3)\\[2pt]
     \frac{2}{k+2}\quad(k\geq4)
   \end{array}$
   \ $\big(\frac12\big)$
& \cite{Fredenhagen:2012rb,Brunner:2000wx} \\[3pt]

Compact boson $\otimes$ MM
& $1+c_m$ \ $\big(\frac32\big)$
& $\Delta_\sigma>0$
& $\min(4\Delta_\sigma,2)$
& \cite{Ginsparg:1987eb,DiFrancesco:1997nk} \\

\bottomrule
\end{tabular}
\caption{
Known unitary CFTs with $c>1$, ordered by the smallest central charge in each family 
%(the values in parentheses are those of the first member; an entry with only a value in parentheses is given for the first member only). 
(the first member in parentheses).
%Except in the last row, the first listed $\sigma$ is the lightest scalar Virasoro primary that does not appear in its own OPE. 
%Further entries give additional operators with this property. 
%In every row, $\eps$ is the lightest scalar in the corresponding $\sigma\times\sigma$ OPE.
Here $\sigma$ and $\eps$ are scalar Virasoro primaries with an OPE of the form~\eqref{eq:sigma-ope}: $\sigma$ does not appear in its own OPE, and $\eps$ is the lightest scalar in $\sigma\times\sigma$.
MM stands for minimal models and $c_m=1-6/[m(m+1)]$ with $m\geq3$.
In the first row, besides the lightest pair, we list the two pairs at the tips of the needles of Figure~\ref{fig:z2-needle}.
The one in bold, $(1/2,8/5)$, comes from the twisted sector of the charge-conjugation orbifold of the $\Z_5$ parafermion, while $(4/7,12/7)$ belongs to the parafermion itself and survives in the orbifold.
In the last row, $\sigma$ is a combination of vertex operators in the compact boson, and $\Delta_\sigma$ varies continuously with the radius.
%Three of the central charges used in Figure~\ref{fig:z2} are realised by theories in the table: $c=8/7$ ($\mathbb{Z}_5$ parafermion and its orbifold), $c=7/5$ ($\mathbb{Z}_8$ parafermion and the $m=10$ $\mathcal{N}=1$ minimal model) and $c=12/5$ ($\mathrm{SU}(2)_8$, the $k=8$ $\mathcal{N}=2$ MM and the $(k,l)=(18,10)$ coset).
}
\label{tab:known-cfts}
\end{table}

\subsection{Numerical results}

We consider a unitary CFT with $c>1$ containing a scalar $\sigma$ that does not appear in its own OPE, as is the case if $\sigma$ is odd under a $\mathbb{Z}_2$ symmetry:
\begin{equation}\label{eq:sigma-ope}
    \sigma\times\sigma \sim \id + \eps + \dots \ec
\end{equation}
where $\eps$ is the lightest scalar Virasoro primary in the OPE. We recall that here, as in the rest of this work, every operator we refer to ($\sigma$, $\eps$, and so on) is a Virasoro primary, and spectra and gaps are always stated in terms of Virasoro primaries. In theories with an extended chiral algebra, such as current, $W$, or superconformal algebras, a representation of the extended algebra generally decomposes into several Virasoro representations, whose primaries each count as a separate operator. From the crossing equation of $\braket{\sigma\sigma\sigma\sigma}$ we derive an upper bound on $\Delta_\eps$ as a function of $\Delta_\sigma$ at fixed $c$. We assume that all scalars in $\sigma\times\sigma$ other than the identity have $\Delta\geq\Delta_\eps$, while spinning operators only satisfy unitarity, $\Delta\geq\ell$. The stress tensor is part of the identity Virasoro block, so the central charge enters only through the blocks. A theory at central charge $c$ with a pair $(\Delta_\sigma,\Delta_\eps)$ above the bound cannot exist. If the bound collapses over a whole range of $\Delta_\sigma$, no CFT with such a $\sigma$ exists at that $c$. Table~\ref{tab:known-cfts} lists the known CFTs with $c>1$ closest to $c=1$, with the dimensions of $\sigma$ and $\eps$ to compare with the bounds.

\begin{figure}[t]
  \centering
  \includegraphics[width=0.7\textwidth]{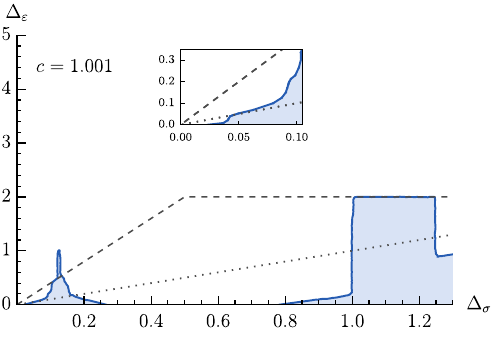}
    \caption{Upper bound on the dimension $\Delta_\eps$ of the lightest scalar in the $\sigma\times\sigma$ OPE, as a
  function of the dimension of a $\mathbb{Z}_2$-odd scalar $\sigma$, from $\braket{\sigma\sigma\sigma\sigma}$ with
  Virasoro blocks for $c=1.001$. The numerical parameters are in Table~\ref{tab:z2-parameters}. The dashed line is the compact boson at $c=1$,
  $\Delta_\eps=\min(4\Delta_\sigma,2)$, while the dotted line is $\Delta_\eps=\Delta_\sigma$.}\label{fig:z2-c1p001}
\end{figure}

\begin{figure}[t]
  \centering
  \begin{subfigure}{0.49\textwidth}
    \includegraphics[width=\textwidth]{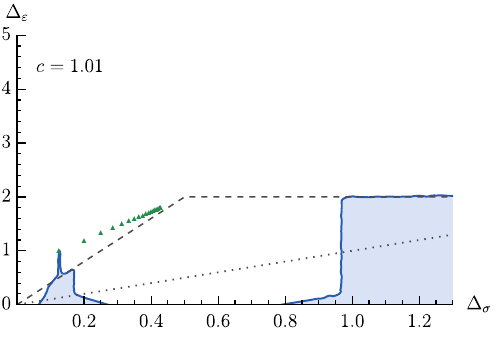}
    \caption{}\label{fig:z2-c1p01}
  \end{subfigure}
  \begin{subfigure}{0.49\textwidth}
    \includegraphics[width=\textwidth]{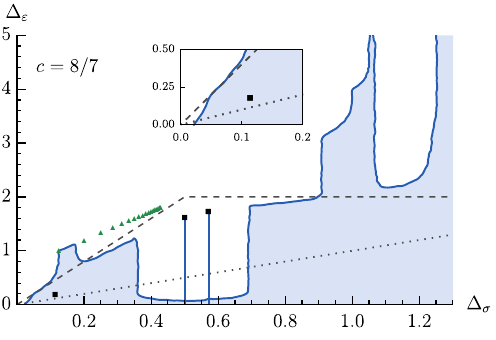}
    \caption{}\label{fig:z2-c8o7}
  \end{subfigure}\\[4pt]
  \begin{subfigure}{0.49\textwidth}
    \includegraphics[width=\textwidth]{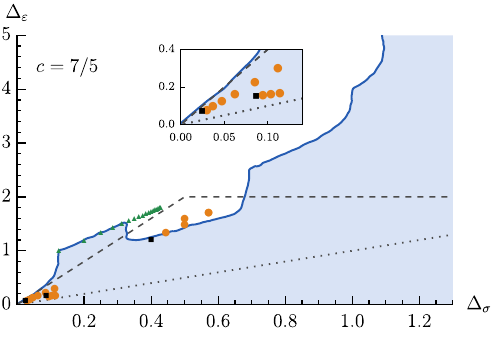}
    \caption{}\label{fig:z2-c7o5}
  \end{subfigure}
  \begin{subfigure}{0.49\textwidth}
    \includegraphics[width=\textwidth]{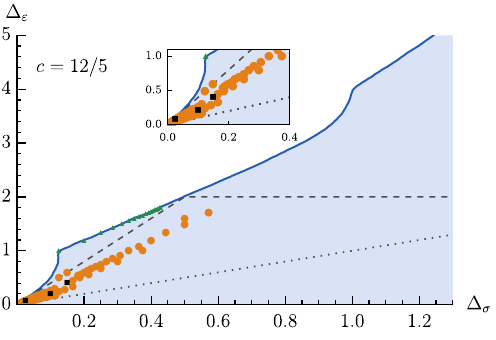}
    \caption{}\label{fig:z2-c12o5}
  \end{subfigure}
  \caption{Upper bound on the dimension $\Delta_\eps$ of the lightest scalar in the $\sigma\times\sigma$ OPE, as a
  function of the dimension of a $\mathbb{Z}_2$-odd scalar $\sigma$, from $\braket{\sigma\sigma\sigma\sigma}$ with
  Virasoro blocks at different values of the central charge. 
  The numerical parameters are in Table~\ref{tab:z2-parameters}. 
  The dashed line is the compact boson at $c=1$,
  $\Delta_\eps=\min(4\Delta_\sigma,2)$, while the dotted line is $\Delta_\eps=\Delta_\sigma$. We also report known theories from
  Table~\ref{tab:known-cfts}. 
  Black squares: theories at the central charge of the panel ($\mathbb{Z}_5$ parafermion and
  its orbifold in (b), see also Figure~\ref{fig:z2-needle}; $\mathbb{Z}_8$ parafermion and the $m=10$ $\mathcal{N}=1$ minimal model in (c);
  $\mathrm{SU}(2)_8$, the $k=8$ $\mathcal{N}=2$ minimal model and the $(k,l)=(18,10)$ coset in (d)).
  Orange disks: theories at smaller central charge ($\mathbb{Z}_k$ parafermions and the $\mathbb{Z}_5$ orbifold,
  $\mathcal{N}=1$ and $\mathcal{N}=2$ minimal models, $\mathrm{SU}(2)_k$ and the $\mathrm{SU}(2)$ cosets up to $k,m=20$). 
  Green triangles: the unitary minimal models
  $\mathcal{M}(m+1,m)$, $c<1$, with $\sigma=\phi_{1,2}$ and $\eps=\phi_{1,3}$, up to $m=20$
  (Appendix~\ref{app:branching}).}
  \label{fig:z2}
\end{figure}

Our results are shown in Figure~\ref{fig:z2-c1p001} for $c=1.001$ and in Figure~\ref{fig:z2} for $c=1.01$, $8/7$, $7/5$ and $12/5$. Close to $c=1$, the bound has a striking structure. At $c=1.001$ (Figure~\ref{fig:z2-c1p001}), it lies well below the compact boson line $\Delta_\eps=4\Delta_\sigma$ for $\Delta_\sigma\lesssim0.1$, and reaches it only at the base of a narrow spike at $\Delta_\sigma=1/8$ (roughly $0.11\lesssim\Delta_\sigma\lesssim0.13$), where it rises to $\Delta_\eps\approx1$. Right after the spike, the bound collapses. It drops below $0.15$ for $0.17\lesssim\Delta_\sigma\lesssim0.96$, is practically zero for $0.3\lesssim\Delta_\sigma\lesssim0.8$, and only at $\Delta_\sigma\approx1$ does it jump back to the compact boson value $\Delta_\eps=2$. This plateau extends to approximately $\Delta_\sigma=1.25$, where the bound drops below the diagonal $\Delta_\eps=\Delta_\sigma$ again. The plot therefore suggests that, in the range where the bound essentially vanishes, there are no CFTs with this value of $c$: the OPE of such a $\sigma$ would have to contain a scalar of almost zero dimension.

The spike has a precise origin, explained in Appendix~\ref{app:at-point}. At $c=1$, the twist fields of the orbifold branch of the compact boson (Appendix~\ref{app:orbifold}) have $\Delta_\sigma=1/8$ for every radius, while the dimensions of the operators in their OPE vary continuously~\cite{Ginsparg:1987eb}; the largest gap along the branch, $\Delta_\eps=1$, is reached at the point of two decoupled Ising models. A theory at $c=1$ is of course not a solution of crossing at $c=1+\epsilon$, and increments as small as $\epsilon=0.001$ or $0.01$ are not covered by the ``spectator'' argument of Appendix~\ref{app:branching}, which applies only to increments larger than one or equal to the central charge of a unitary minimal model. Nevertheless, $\Delta_\sigma=1/8$ is special: it is the only external dimension at which the Virasoro blocks are continuous at $c=1$, and the four-point functions of the whole orbifold branch expand with positive coefficients in blocks at $c=1+\epsilon$, as we verify order by order in Appendix~\ref{app:at-point} for $\epsilon=0.01$ and, at the point of two decoupled Ising models, for $\epsilon=0.001$. The bound at $\Delta_\sigma=1/8$ must therefore stay above $\Delta_\eps=1$, which is exactly the height of the spike. The correlators of the circle branch have instead no positive expansion in blocks at $c=1.01$, so nothing protects the compact-boson line away from $\Delta_\sigma=1/8$.

\begin{figure}[t]
  \centering
  \includegraphics[width=0.6\textwidth]{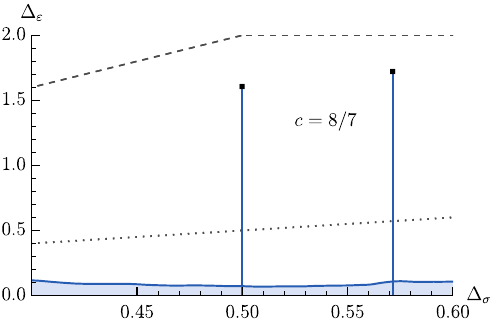}
  \caption{Zoom on the two needles of the bound at $c=8/7$ (Figure~\ref{fig:z2-c8o7}), at $\Delta_\sigma=1/2$ and
  $\Delta_\sigma=4/7$, with the allowed region shaded as in Figure~\ref{fig:z2-c8o7}. On the line $\Delta_\sigma=1/2$
  the bound reaches $1.59<\Delta_\eps<1.61$, and on the line $\Delta_\sigma=4/7$ it reaches $12/7\leq\Delta_\eps<1.73$.
  The needles are much narrower than the resolution of the figure: at $\Delta_\eps=0.1$ the points at
  $|\Delta_\sigma-1/2|=5\times10^{-4}$ are already excluded. The black squares are the two pairs of the
  $\Z_5$ parafermion and its orbifold, $(\Delta_\sigma,\Delta_\eps)=(1/2,8/5)$ and $(4/7,12/7)$
  (Table~\ref{tab:known-cfts}), which sit at the tips of the two needles.}
  \label{fig:z2-needle}
\end{figure}

Passing to larger central charges, the picture changes gradually. The spike broadens, and the allowed regions on the two sides of the collapsed window grow towards each other. Already at $c=1.01$ (Figure~\ref{fig:z2-c1p01}), the bound follows the compact boson line from the spike up to $\Delta_\sigma\approx0.16$ before collapsing. It returns to $\Delta_\eps=2$ at $\Delta_\sigma\approx0.97$, while the interval where it lies below $\Delta_\eps=\Delta_\sigma$ shrinks to $0.18\lesssim\Delta_\sigma\lesssim0.97$. At $c=8/7$ (Figure~\ref{fig:z2-c8o7}) only a small interval, $0.34\lesssim\Delta_\sigma\lesssim0.70$, still has a collapsed bound.
Inside this interval, the bound has two very narrow needles, at $\Delta_\sigma=1/2$ and $\Delta_\sigma=4/7$, which reach $\Delta_\eps\approx1.60$ and $\Delta_\eps\approx1.72$ respectively (Figure~\ref{fig:z2-needle}). Both are realised by theories at exactly this central charge (Table~\ref{tab:known-cfts}). The pair $(4/7,12/7)$ belongs to the $\mathbb{Z}_5$ parafermions and survives in their charge-conjugation orbifold. The pair $(1/2,8/5)$ comes from a twist field of the orbifold. These needles are remarkably narrow. On either side, the bound is smooth and close to zero. As $c$ increases further, the window shrinks, and it closes somewhere between $c=8/7$ and $c=7/5$ (Figure~\ref{fig:z2-c7o5}). This is also the range in which, after the $\mathbb{Z}_5$ parafermions and their orbifold at $c=8/7$, the next known theories appear (Table~\ref{tab:known-cfts}). Beyond this point, the bound takes a shape very similar to the one obtained with global conformal blocks~\cite{Rychkov:2009ij,El-Showk:2012vjm}.

The curve at $c=12/5$ (Figure~\ref{fig:z2-c12o5}) indeed already has the shape of the global bound, including the kink at $(1/8,1)$, which corresponds to the Ising model in the global analysis. This is somewhat expected. A Virasoro block is a sum over the global conformal blocks of the quasi-primaries in the Verma module, and as $c\to\infty$ at fixed weights only the first of them survives. This is manifest in the original form of Zamolodchikov's recursion~\cite{Zamolodchikov:1984eqp}, which expands the chiral block around its large-$c$ limit,
\begin{equation}
    V_h(\z) = g_h(\z) + \sum_{m\geq1 , n\geq2} \frac{R'_{m,n}(h)}{c-c_{m,n}(h)}\, V_{h+mn}(\z)\Big|_{c=c_{m,n}(h)} \ec
\end{equation}
where $g_h(\z)=\z^h\hypgeo{2}{1}(h,h,2h;\z)$ is the global block and $R'_{m,n}(h)$ is the residue coefficient at $c=c_{m,n}(h)$, with its dependence on the fixed external weight suppressed. Every correction comes with a pole in $c$ and vanishes as $c\to\infty$. The recursion~\eqref{eq:H-recursion} that we use in practice is equivalent, but there the large-$c$ limit is less transparent, since the dependence on $c$ is shared between $H$ and the prefactor in~\eqref{eq:chiral-prefactor}. This suggests that the bound approaches the one obtained with global conformal blocks as $c$ increases~\cite{Rychkov:2009ij}.

More generally, Appendix~\ref{app:branching} shows that a Virasoro block at central charge $c$ is a sum with non-negative coefficients of blocks at $c'=c+s$ whenever $s\geq1$ or $s$ is the central charge of a unitary minimal model. The region allowed at $c'$ then contains every solution of crossing at $c$ with $\Delta_\eps\leq4$. This is why the Ising model, all the unitary minimal models, and the theories at $c=1$ must lie inside the region allowed at $c=12/5$. The Ising point and the indicated minimal-model points additionally appear to saturate the bound. No such statement holds when $c'-c$ is small, and the collapse of the bound at $c=1.001$ and $1.01$ below the compact-boson line, which is realised at $c=1$, is possible precisely for this reason.

We can now come back to the conjecture. Our results do not prove it, but they exclude a well-defined class of theories close to $c=1$: those whose lightest scalar $\sigma$ does not appear in its own OPE, as happens for instance when $\sigma$ is odd under a $\mathbb{Z}_2$ symmetry. In such a theory every scalar in $\sigma\times\sigma$ is at least as heavy as $\sigma$, so $\Delta_\eps\geq\Delta_\sigma$, and the theory is excluded whenever the bound lies below the line $\Delta_\eps=\Delta_\sigma$. This happens in the following ranges:
\begin{itemize}
\item at $c=1.001$, for $0.17\lesssim\Delta_\sigma\lesssim1.00$, which is essentially the whole collapsed window;
\item at $c=1.01$, for $0.18\lesssim\Delta_\sigma\lesssim0.97$, almost the same interval as at $c=1.001$;
\item at $c=8/7$, for $0.34\lesssim\Delta_\sigma\lesssim0.70$, apart from the two needles at $\Delta_\sigma=1/2$ and $\Delta_\sigma=4/7$.
\end{itemize}
At $c=7/5$ and $c=12/5$ the bound never crosses the line, and no theory of this class is excluded.

If $\sigma$ is also the lightest primary of any spin, so that $\Delta\ped{gap}=\Delta_\sigma$, this can be combined with the modular bound $\Delta\ped{gap}\leq c/6+1/3$, which gives $\Delta_\sigma\lesssim0.50$ at $c=1.001$ and $\Delta_\sigma\lesssim0.52$ at $c=8/7$. Such a theory can then exist only if $\Delta_\sigma\lesssim0.17$ at $c=1.001$, or, at $c=8/7$, if $\Delta_\sigma\lesssim0.34$ or $\Delta_\sigma=1/2$.
These exclusions are rigorous at the derivative order we used, up to the usual controlled approximations of the numerical bootstrap, and agree with the results of~\cite{Nakayama:2026tba}.

It is instructive to compare this result with the modular bootstrap. The authors of~\cite{Benjamin:2026lbj} searched for truncated modular-invariant partition functions with integer degeneracies, and found none for $1<c<1.06$ with $0.3<\Delta\ped{gap}<0.5$, just below the bound $\Delta\ped{gap}\leq c/6+1/3$. Since this region fills up once the degeneracies are allowed to be non-integer, they attributed it to integrality. Under the assumptions above, our results exclude the same region at $c=1.001$, and in fact a larger one, without any use of integrality. In this sense the two approaches are complementary: the modular bootstrap constrains the full spectrum, but only through the torus partition function, while crossing symmetry excludes a larger region near $c=1$, at the price of requiring the lightest operator to be a scalar absent from its own OPE.

\section{Conclusions} 

In this work we developed a numerical bootstrap for two-dimensional CFTs with $c>1$ that exploits the full Virasoro symmetry. The blocks are computed with Zamolodchikov's recursion and brought, through optimal polynomial interpolation, to the form required by semi-definite programming. The restriction to $c>1$ is intrinsic: only there do the poles of the blocks leave the positive real axis. As a first test, we isolated a small island around three decoupled Ising models. 
For three 3-state Potts models coupled through their energy operators, the mixed system of the $\phi$ and $z$ correlators isolates the decoupled theory in an island when the gaps follow its spectrum. Relaxing the gaps where perturbation theory predicts negative anomalous dimensions, and raising the leading scalar-singlet gap, gives instead a peninsula at $c=2.4$ that contains the estimated scaling dimensions of the coupled fixed point and excludes the decoupled theory.

In parallel with these studies, we explored the space of CFTs with central charge just above one. There, the bound on the lightest scalar in the OPE of two identical scalars collapses over a wide range of external dimensions, and a theory whose lightest scalar is absent from its own OPE is excluded in that range. What survives are remnants of the $c=1$ theories: a spike at $\Delta_\sigma=1/8$, inherited from the orbifold branch, and a plateau at the marginal dimension $\Delta_\eps=2$. The excluded window shrinks as $c$ grows and closes between $c=8/7$ and $c=7/5$. At $c=8/7$, however, the collapsed window contains two extremely narrow needles, much thinner than the resolution of the plots, whose tips sit exactly at operators of the $\mathbb{Z}_5$ parafermions and of their charge-conjugation orbifold, two known theories at precisely this central charge. Furthermore, some of the features observed in the plots were explained by exact properties of the blocks: a block at central charge $c$ is a positive sum of blocks at $c+s$ whenever $s\geq1$ or $s$ is the central charge of a unitary minimal model, so the regions allowed at different central charges are nested (for $\Delta_\eps\leq4$).

The code presented here can be used to study conjectured two-dimensional CFTs systematically, and the analyses of Sections~\ref{sec:ising} and~\ref{sec:potts} indicate that Virasoro symmetry has a non-negligible effect on the bounds. One reason is that the central charge can be injected exactly: with global blocks it enters only through the OPE coefficient of the stress tensor, while here it enters every block. The price is one additional parameter on which the bounds depend. When $c$ is not known exactly, as for the coupled Potts models with $c\simeq2.377$~\cite{Dotsenko:1998gyp}, one has to scan over it, or navigate the space of parameters with the navigator function of~\cite{Reehorst:2021ykw}. A second reason is that organising the spectrum in Virasoro multiplets makes gap assumptions more efficient. With global blocks, the quasi-primary descendants of each Virasoro primary must be allowed in the spectrum, which caps the gaps one can impose; with Virasoro blocks, their contributions are fixed, and a gap on primaries carries much more information. A systematic comparison between Virasoro and global blocks will be the subject of a future publication. We expect the difference to be most pronounced at small $c$, where the two kinds of blocks differ most, and for mixed correlators.

The bounds on the coupled Potts models are interesting but very sensitive to the gaps. Here we chose conservative assumptions. Stronger, independently justified spectral assumptions could produce an island for the coupled theory or rule out its proposed CFT data. To justify them, one needs more precise estimates of its low-lying spectrum, from microscopic methods such as transfer matrices, quantum spin-chain simulations and Monte Carlo, which we are currently pursuing~\cite{Kousvos:morePotts}. Similar studies apply to other candidate irrational CFTs, such as coupled minimal models~\cite{Antunes:2022vtb,Antunes:2024mfb,Antunes:2025erb} and theories with non-invertible symmetries~\cite{Antunes:2025huk,Ferragatta:2026ejs}. External spinning operators, in particular higher-spin currents, would make it possible to probe integrable theories. Another direction is the mixed system of the annulus partition function, bulk two-point functions in the presence of a boundary, and bulk four-point functions developed in~\cite{Meineri:2025qwz}. Incorporating Virasoro blocks throughout would combine Virasoro characters and blocks in the same bootstrap problem.

Virasoro blocks also appear beyond two-dimensional CFTs. In four-dimensional \mbox{$\mathcal{N}\geq2$} superconformal theories, a protected sector of the correlators is captured by a $2d$ chiral algebra~\cite{Beem:2013sza}, whose Virasoro symmetry enters the four-dimensional crossing equations~\cite{Beem:2014zpa}. Similarly, in celestial holography, scattering amplitudes in four-dimensional flat space are recast as correlators on the celestial sphere, which display the structure of a two-dimensional CFT~\cite{Strominger:2017zoo}. These applications lie outside the unitary two-dimensional setup studied here: the associated chiral algebra has negative central charge, while celestial correlators do not obey the usual two-dimensional unitarity conditions. The blocks, however, are given by the same recursion at any central charge. Extending the block approximation to the relevant parameters could be useful in both settings, but applying numerical bootstrap methods would also require an appropriate treatment of their spectra and positivity constraints.

We conclude with the CFT desert. With a single correlator we excluded, just above $c=1$, every theory whose lightest scalar is absent from its own OPE and lies in a wide range of dimensions, which supports the conjecture that no unitary and compact CFT exists in this region. Several strategies could go further. Dropping the assumption that $\sigma$ is absent from its own OPE, and including its block with a free coefficient, would make the exclusion apply to theories without a $\mathbb{Z}_2$ symmetry. Mixed correlators of $\sigma$ and $\eps$, which in higher dimensions are far more constraining than a single one, could probe the regions that survive close to the $c=1$ theories. A gap above $\Delta=2$ in the spin-two Virasoro-primary sector would remove solutions containing additional conserved spin-two currents and could test the proposed origin of the spike. An example is the current $T_1-T_2$ of two decoupled Ising models at the top of the spike. Combining crossing with modular invariance in a single system would use the full spectrum and crossing at the same time.

Forty years after Belavin, Polyakov and Zamolodchikov, the landscape of two-dimensional CFTs with $c>1$ remains largely uncharted, and we hope that the Virasoro bootstrap will help to map it.

\section*{Acknowledgements}

We thank Yu Nakayama for suggesting the investigation of theories with central charge close to 1 during the workshop ``Progress of Theoretical Bootstrap'' at YITP, Kyoto, in 2025. We thank Ant\'onio Antunes, Petr Kravchuk, Johan Henriksson, Jiaxin Qiao, Slava Rychkov and Ning Su for useful discussions.
We thank the organisers of the conference Bootstrap 2025 at Instituto Principia in São Paulo and the conference Bootstrap 2026 at City St George's, University of London and Imperial College London for the hospitality and for providing a stimulating
environment where part of this work was carried out.

SRK, AP have been partially supported by the INFN ``Iniziativa Specifica'' STEFI. AV has been partially supported by the INFN ``Iniziativa Specifica'' SFT. SRK has received support from the Marie Sk\l{}odowska-Curie Action (MSCA) High energy Intelligence (HORIZON-MSCA-2023-SE-01-101182937-HeI). SRK is supported by the Italian Ministry of University and Research (MUR) under the FIS grant BootBeyond (CUP: D53C24005470001). 
AP is supported by the UK Engineering and Physical Sciences Research Council under grant EP/Z000106/1.
AV is supported by the Italian Ministry of University and Research (MUR) through the FIS Advanced grant QBOOT (contract FIS-2024-005507, CUP: I53C25002240001).

The computations presented here were conducted on the SISSA HPC cluster Ulysses, the INFN Pisa HPC cluster Theocluster Zefiro, the Asteria HPC cluster (affectionately referred to as gattaiola) of the University of Torino, and the facilities provided by the Imperial College Research
Computing Service (\url{http://doi.org/10.14469/hpc/2232}).

We acknowledge the use of Claude (Anthropic) and ChatGPT (OpenAI) as assistive tools for coding, for polishing the manuscript, and for discussing ideas.
The implementation of the Virasoro blocks and the bootstrap setup is based on human-written code (pre-AI) by the authors.
It was later polished and slightly improved by AI tools.

\appendix

\section{Known Virasoro blocks}\label{app:known-blocks}
We collect several closed-form expressions for chiral Virasoro blocks used to check our implementation and the normalisation $V_h(z)\sim z^h$.

For a correlator containing one level-2 degenerate primary $\phi_{2,1}$ of conformal weight $h_{2} = -\tfrac34 b^{2}-\tfrac12$ and three generic Virasoro primaries of weights $h_{1},h_{3},h_{4}$, there are two possible exchanged weights \( h_{\pm} = \lambda_{1,1}^{2} - \left(\lambda_{1}\pm\tfrac{b}{2}\right)^{2} \), whose corresponding chiral blocks are~\cite{Ribault:2014hia}
\begin{equation}
    \begin{aligned}
        V_{h_{-}}(\z) &= \z^{h_{-}}(1-\z)^{\,b(\lambda_{1,1}+\lambda_{3})} \\
        &\quad\times \hypgeo{2}{1}\!\left(\tfrac12+b(\lambda_{1}+\lambda_{3}+\lambda_{4}),\,\tfrac12+b(\lambda_{1}+\lambda_{3}-\lambda_{4});\,1+2b\lambda_{1};\,\z\right) \ec \\
        V_{h_{+}}(\z) &= \z^{h_{+}}(1-\z)^{\,b(\lambda_{1,1}+\lambda_{3})} \\
        &\quad\times \hypgeo{2}{1}\!\left(\tfrac12+b(-\lambda_{1}+\lambda_{3}+\lambda_{4}),\,\tfrac12+b(-\lambda_{1}+\lambda_{3}-\lambda_{4});\,1-2b\lambda_{1};\,\z\right) \ed
    \end{aligned}
\end{equation}
with $b$ as in~\eqref{eq:b-branch} and $\lambda_i$ and $\lambda_{1,1}$ given by~\eqref{eq:lambda-parameters}.

A correlator of four level-3 degenerate primaries \( \phi_{1,3} \) of conformal weight $h_{1}=h_{2}=h_{3}=h_{4}=\tfrac35$ at central charge $c = 7/10$ admits exchanges of weights $h = 0, 3/5$, and the corresponding chiral blocks are~\cite{Gu:2023plq}:
\begin{equation}
    \begin{aligned}
        V_{0}(\z) &= (1-\z)^{-3/5}(\z^{2}-\z+1)\,\hypgeo{2}{1}\!\left(\tfrac15,\tfrac45;\tfrac25;\z\right) \ec \\
        V_{3/5}(\z) &= \z^{3/5}(1-\z)^{-3/5}(\z^{2}-\z+1)\,\hypgeo{2}{1}\!\left(\tfrac45,\tfrac75;\tfrac85;\z\right) \ed
    \end{aligned}
\end{equation}

A correlator of four level-4 degenerate primaries \( \phi_{2,2} \) of conformal weight $h_{1}=h_{2}=h_{3}=h_{4}=-\tfrac34\left(b^{2}+b^{-2}\right)-\tfrac32$ admits, at generic central charge, exchanges of four conformal weights
\begin{equation}
    h=0 \ec \quad
    h_{3,1} = -2b^{2}-1 \ec \quad
    h_{1,3} = -2b^{-2}-1 \ec \quad
    h_{3,3} = -2b^{2}-2b^{-2}-4 \ed
\end{equation}
At minimal-model central charges, additional fusion constraints can remove channels. With the eight auxiliary functions
\begin{equation}
    \begin{aligned}
        s_{1} &= \hypgeo{2}{1}\!\left(1+b^{2},\,3+3b^{2};\,2+2b^{2};\z\right) \ec &
        s_{2} &= \hypgeo{2}{1}\!\left(1+b^{2},\,3+3b^{2};\,3+2b^{2};\z\right) \ec \\
        s_{3} &= \hypgeo{2}{1}\!\left(2+b^{2},\,-b^{2};\,-2b^{2};\z\right) \ec &
        s_{4} &= \hypgeo{2}{1}\!\left(1+b^{2},\,-1-b^{2};\,-1-2b^{2};\z\right) \ec \\
        t_{1} &= \hypgeo{2}{1}\!\left(4+3b^{-2},\,2+b^{-2};\,3+2b^{-2};\z\right) \ec &
        t_{2} &= \hypgeo{2}{1}\!\left(4+3b^{-2},\,2+b^{-2};\,4+2b^{-2};\z\right) \ec \\
        t_{3} &= \hypgeo{2}{1}\!\left(-b^{-2},\,2+b^{-2};\,-1-2b^{-2};\z\right) \ec &
        t_{4} &= \hypgeo{2}{1}\!\left(-1-b^{-2},\,1+b^{-2};\,-2-2b^{-2};\z\right) \ed
    \end{aligned}
\end{equation}
the blocks are~\cite{Henriksson:2025kws}
\begin{equation}
    \begin{aligned}
        V_{0}(\z) &= (1-\z)^{-\Delta_{\phi}}\left[(1-\z)\,s_{1} t_{1} - \frac{1+b^{2}}{2(2+3b^{2})}\,\z\, s_{2} t_{2}\right] \ec \\
        V_{h_{3,1}}(\z) &= \frac{\z^{h_{3,1}}(1-\z)^{-\Delta_{\phi}}}{1+b^{2}}\left[(1-\z)(2+3b^{2})\,s_{3} t_{1} - (1+2b^{2})\,s_{4} t_{2}\right] \ec \\
        V_{h_{1,3}}(\z) &= \frac{2(2+b^{2})(3+2b^{2})}{3(1+b^{2})^{2}}\,\z^{h_{1,3}-1}(1-\z)^{-\Delta_{\phi}}\left[(\z-1)\,s_{1} t_{3} + s_{2} t_{4}\right] \ec \\
        V_{h_{3,3}}(\z) &= \z^{h_{3,3}}(1-\z)^{-\Delta_{\phi}}\left[-\frac{1+b^{2}}{2(1+2b^{2})}\,\z(1-\z)\,s_{3} t_{3} + s_{4} t_{4}\right] \ed
    \end{aligned}
\end{equation}
where $\Delta_\phi = -\tfrac32\left(b^{2}+b^{-2}\right)-3$.

For four twist fields with $h_{1}=h_{2}=h_{3}=h_{4}=\tfrac{1}{16}$ at $c = 1$, and for non-degenerate exchanged weight $h$, the chiral block is the Ashkin--Teller block~\cite{Zamolodchikov:1986gt,Nemkov:2021huu},
\begin{equation}
    V_{h}(\z) = \bigl(16\,q(\z)\bigr)^{h}\,(1-\z)^{-1/8}\,\vartheta_{3}\bigl(q(\z)\bigr)^{-1} \ed
\end{equation}
with $q$ the elliptic nome in~\eqref{eq:elliptic-nome}.
At degenerate weights, this expression gives the \(U(1)\) block, which decomposes into irreducible Virasoro blocks.

\section{The circle and orbifold branches at $c=1$}\label{app:orbifold}

This appendix collects the facts about the $c=1$ theories used in Section~\ref{sec:cft-desert} and in Appendix~\ref{app:at-point}.
See~\cite{Ginsparg:1987eb} for a complete account.
The compact boson $X\sim X+2\pi R$ has Virasoro primaries $V_{n,w}$ labelled by the momentum $n$ and the winding $w$, with weights
\begin{equation}\label{eq:boson-weights}
    (h,\bar h)=\left(\frac{p_L^2}{2},\frac{p_R^2}{2}\right)\ec\qquad p_{L,R}=\frac nR\pm\frac{wR}{2}\ec
\end{equation}
in conventions where $R\to2/R$ exchanges $n$ and $w$ and the self-dual radius is $R=\sqrt2$. For generic $p_{L,R}$, the Fock module built on $V_{n,w}$ has no null states and is a single Verma module.
When \(p_L^2/2=j_0^2\), with \(j_0\in\frac12\mathbb Z_{\geq0}\), the left-moving Fock module decomposes into irreducible Virasoro modules of weights \((j_0+k)^2\), \(k=0,1,\ldots\), whose highest-weight states are the discrete states (and analogously for the right-moving sector)~\cite{Gaberdiel:2001zq}.
This is the circle branch of the $c=1$ moduli space.

The operator of Section~\ref{sec:cft-desert} is, on this branch, a vertex operator $\sigma=\sqrt2\cos(pX)$ with $p=n/R$ and $\Delta_\sigma=p^2$.
It is odd under the shift $X\to X+\pi/p$, a $\mathbb Z_2$ subgroup of the momentum $U(1)$, and $\Delta_\sigma$ takes every positive value as $R$ varies. By charge conservation, $\sigma\times\sigma$ contains the neutral sector and the operators of charge $\pm2p$, the lightest of which is $\cos(2pX)$ with $\Delta=4\Delta_\sigma$.
The lightest neutral scalar primary is the marginal operator $\partial X\bar\partial X$, with $\Delta=2$ and $\lambda^2_{\sigma\sigma\,\partial X\bar\partial X}=\Delta_\sigma^2$. Hence $\Delta_\eps=\min(4\Delta_\sigma,2)$, which is reported as a dashed line in Figure~\ref{fig:z2}. The four-point function is a free-field computation and reads, in the conformal frame,
\begin{equation}\label{eq:boson-4pt}
    \begin{aligned}
        (z\bar{z})^{2h_{\sigma}}\braket{\sigma(\infty)\sigma(1)\sigma(z,\bar{z})\sigma(0)}
        &=\frac12\left[(1-\z)^{2h_\sigma}(1-\zb)^{2h_\sigma}+(1-\z)^{-2h_\sigma}(1-\zb)^{-2h_\sigma}\right]\\
        &\quad +\frac12\,\z^{4h_\sigma}(1-\z)^{-2h_\sigma}\,\zb^{4h_\sigma}(1-\zb)^{-2h_\sigma}\ed
    \end{aligned}
\end{equation}
The last term is the contribution of the $U(1)$ modules of charge $\pm2p$, whose chiral factor $\z^{4h_\sigma}(1-\z)^{-2h_\sigma}$ is, for generic $p$, the $c=1$ Virasoro block of weight $4h_\sigma$. 
The first two terms are the neutral sector: the identity block together with the discrete states with $j+j'$ even. These are the functions expanded in blocks at $c=1.01$ in Appendix~\ref{app:at-point}.

The orbifold branch is obtained by gauging the symmetry $X\to-X$. Its untwisted sector contains the even combinations $V_{n,w}+V_{-n,-w}$ and the oscillator states that are even under $X\to-X$. Its twisted sector is built on the two fixed points $X=0$ and $X=\pi R$, around which $X$ is antiperiodic: the corresponding ground states are the twist fields $\sigma_1$ and $\sigma_2$, with $h=\bar h=1/16$, the zero-point energy of an antiperiodic boson. Their dimension does not depend on $R$, which is why the whole branch sits at $\Delta_\sigma=1/8$ in Figure~\ref{fig:z2}, while the dimensions of the operators in $\sigma\times\sigma$ vary with the radius.

The OPE of a twist field with itself contains the untwisted operators with even momentum, and the four-point function was computed in~\cite{Zamolodchikov:1986gt} (see also~\cite{Dixon:1986qv}),
\begin{equation}\label{eq:twist-4pt}
    \braket{\sigma_1(\infty)\sigma_1(1)\sigma_1(\z,\zb)\sigma_1(0)}=\frac{\abs{\vartheta_3(q)}^{-2}}{\abs{\z(1-\z)}^{1/4}}\sum_{n,w\in\mathbb Z}q^{\frac12\left(\frac nr+wr\right)^2}\bar q^{\frac12\left(\frac nr-wr\right)^2}\ec\qquad r=\frac R2\ec
\end{equation}
with $q$ the elliptic nome~\eqref{eq:elliptic-nome}. Writing $q^{h}=16^{-h}(16q)^h$ and stripping the prefactor $\abs{\z}^{-1/4}$ of~\eqref{eq:four-point}, each term is $16^{-h-\bar h}$ times the product of the Ashkin--Teller functions $(16q)^h\Phi(\z)$ and $(16\bar q)^{\bar h}\Phi(\zb)$ of Appendix~\ref{app:known-blocks}, with $\Phi=(1-\z)^{-1/8}\vartheta_3(q)^{-1}$: this is the contribution of the whole $U(1)$ module of charge $(p_L,p_R)$, which coincides with a Virasoro block for generic charges and is a positive sum of degenerate Virasoro blocks otherwise. The term $n=w=0$ is the $U(1)$ vacuum module, $\Phi(\z)\Phi(\zb)$, which contains the Virasoro identity block and the even discrete states, the first of which has $(h,\bar h)=(4,0)$.

The lightest scalar in $\sigma_1\times\sigma_1$ is the lightest of the momentum states $n=\pm1$, $w=0$, of dimension $1/r^2$, and of the winding states $n=0$, $w=\pm1$, of dimension $r^2$. Along the orbifold branch, the gap is therefore
\begin{equation}\label{eq:orbifold-gap}
    \Delta_\eps(r)=\min\big(r^{-2},r^2\big)\leq1\ec
\end{equation}
with equality at $r=1$, where the theory is two decoupled Ising models.
At this point, the twist fields are the two spin fields, the degenerate momentum and winding states combine into $\eps_1$, with $\lambda^2_{\sigma\sigma\eps}=4/16=1/4$ from~\eqref{eq:twist-4pt}, and the states $(n,w)=\pm(1,1)$ give the spin-two current $T_1-T_2$ of weight $(2,0)$, with $\lambda^2=2/16^2=1/128$, in agreement with~\eqref{eq:alpha2-vac} for $c=s=1/2$. The point $(\tfrac18,1)$ is thus the top of the orbifold branch, and the whole branch is what lifts to $c=1+\epsilon$ in Appendix~\ref{app:at-point}.

\section{Virasoro blocks at different central charges}\label{app:branching}

The bounds of Section~\ref{sec:cft-desert} depend on the central charge only through the blocks, and bounds at different values of $c$ are not independent. In this appendix we prove one exact property of Virasoro blocks and derive its consequences for the bounds. The property is a positive branching rule: a block at central charge $c$ is a sum, with non-negative coefficients, of blocks at central charge $c'=c+s$, whenever $s$ is the central charge of a unitary vacuum representation of the Virasoro algebra, that is, whenever $s\geq1$ or $s$ is the central charge of a unitary minimal model (Section~\ref{sec:branching-proof}). When both theories exist, this is the familiar statement that a CFT at central charge $c$, tensored with a decoupled spectator theory of central charge $s$, is a CFT at central charge $c'$; the branching rule shows that it remains true when the spectator is only a vacuum representation, and the solution of crossing is only a point allowed by the numerics. It has four consequences.
\begin{itemize}
\item \emph{Nested allowed regions.} A point $(\Delta_\sigma,\Delta_\eps)$ allowed at $c$ with $\Delta_\eps\leq4$ is also allowed at $c+s$, so that $\Delta_\eps^{\max}(c+s,\Delta_\sigma)\geq\min\big(\Delta_\eps^{\max}(c,\Delta_\sigma),4\big)$. The statement holds for the numerical bounds at any derivative order, because a functional that excludes a point at $c+s$ also excludes it at $c$ (Section~\ref{sec:nested}). The threshold at four is the scalar $W\bar W$ built from the second conserved spin-two current $W$ of the product theory, which the spectator always brings along.
\item \emph{Comparison with global blocks.} As $s\to\infty$ the blocks at $c'$ reduce to global blocks. The Virasoro bound at any $c>1$ therefore lies below the bound obtained with global blocks~\cite{Rychkov:2009ij} wherever the latter is below four; this is the behaviour of Figure~\ref{fig:z2-c12o5}.
\item \emph{Theories at lower central charge inside the bounds.} The region allowed at $c'$ contains every solution of crossing at $c'-s$, for every admissible $s$; at $c'=12/5$, for instance, it contains all unitary theories with $c\leq7/5$. The rule guarantees only that these points are allowed, not that they sit on the boundary. That the Ising point is the kink of the $c=12/5$ bound, and that the other minimal models lie along its boundary, is an observation (Section~\ref{sec:comparison}), which parallels the behaviour of the global-block bound~\cite{Rychkov:2009ij,El-Showk:2012vjm}. Conversely, a single-correlator bound at $c'$ cannot tell a genuine theory at $c'$ apart from a theory at lower central charge with a decoupled spectator. All the latter contain the spin-two primary $W$ at $\Delta=2$, so that a twist gap in the spin-two sector removes them.
\item \emph{No constraint for other increments.} For $s$ outside this set, the branching coefficients can be negative. In particular, nothing forces the $c=1$ theories to remain allowed at $c=1+\epsilon$, which is what makes the collapse of the bound and the desert of Section~\ref{sec:cft-desert} possible. The rule does not predict the collapse, nor the spike at $\Delta_\sigma=1/8$, whose origin is different and is explained in Section~\ref{app:at-point}.
\end{itemize}

\subsection{Positive branching of the blocks}\label{sec:branching-proof}

We consider four identical external scalars of weight $h_\sigma=\Delta_\sigma/2$ and make the dependence of the chiral blocks on the central charge explicit, $V_h(\z)\to V^{(c)}_h(\z)$, with $V^{(c)}_h(\z)\sim\z^h$ as in~\eqref{eq:virasoro-block-full}. We denote by $c_m=1-6/(m(m+1))$, $m=3,4,\dots$, the central charges of the unitary minimal models, and by $\mathcal H_{c,h}$ a unitary irreducible representation of the Virasoro algebra of weight $h$ at central charge $c$: for $c>1$ any $h\geq0$, for $c=c_m$ a weight of the Kac table~\cite{Friedan:1983xq}.

\paragraph{Statement.} Let
\begin{equation}\label{eq:spectator-charges}
    c'=c+s\ec\qquad s\in\{c_3,c_4,\dots\}\cup[1,\infty)\ec
\end{equation}
so that a unitary vacuum representation $\mathcal V_s$ of the Virasoro algebra with central charge $s$ exists~\cite{Friedan:1983xq,Goddard:1986ee}. Then
\begin{equation}\label{eq:branching}
    V^{(c)}_h(\z)=\sum_{N=0}^{\infty}\alpha_N\,V^{(c')}_{h+N}(\z)\ec\qquad \alpha_0=1\ec\quad \alpha_N\geq0\ec\quad \alpha_{2k+1}=0\ec
\end{equation}
and the series converges for $|\z|<1$. The coefficients depend on $h$, $h_\sigma$, $c$ and $c'$.

\paragraph{Proof.} The block is a sum over the states of the exchanged representation. Let $\{\ket{n,a}\}$ be an orthonormal basis of $\mathcal H_{c,h}$ at level $n$, and let $A(v)=\braket{\sigma|\sigma(1)|v}/\braket{\sigma|\sigma(1)|h}$ be the normalised three-point amplitude of a state $v$. Then
\begin{equation}\label{eq:state-sum}
    V^{(c)}_h(\z)=\sum_{n\geq0}\z^{h+n}\sum_a\abs{A(\ket{n,a})}^2\ed
\end{equation}
This is the familiar expression of the block through the inverse Gram matrix, written in an orthonormal basis. The two three-point amplitudes coincide because the four external operators are equal. The amplitudes are determined recursively from $A(\ket h)=1$ and the Ward identity 
\begin{equation}\label{eq:ward}
    A(L_{-n}v)=\big(w+(n-1)h_\sigma\big)A(v)\ec\qquad n\geq1\ec
\end{equation}
for a state $v \in \mathcal{H}_{c,h}$ of weight $w$. For the vacuum, $A(L_{-1}\ket0)=0$, consistently with $L_{-1}\ket0=0$.

We now tensor the representation with a ``spectator'' vacuum representation, $\mathcal K=\mathcal H_{c,h}\otimes\mathcal V_s$, and extend the amplitudes as $\widetilde A(v\otimes u)=A(v)\braket{0|u}$. In an orthonormal basis of $\mathcal K$, only the states with $u=\ket0$ have a non-vanishing amplitude, so the state sum~\eqref{eq:state-sum} over $\mathcal K$ still gives $V^{(c)}_h(\z)$. Physically, if both theories exist, this is the four-point function of $\sigma\otimes\id$ in the decoupled product theory of the original theory with a theory of central charge $s$. The point is that $\mathcal K$ is also a unitary representation of the total Virasoro algebra, generated by $L_n=L_n^{(c)}+L_n^{(s)}$, with central charge $c'=c+s$.
Moreover, $\widetilde A$ obeys the Ward identity~\eqref{eq:ward} with respect to the total generators: the spectator term $A(v)\braket{0|L^{(s)}_{-n}u}$ vanishes because $\bra0L^{(s)}_{-n}=0$, and $\braket{0|u}$ vanishes unless $u$ has level zero, so that the weight of $v$ in~\eqref{eq:ward} can be replaced by the total weight of $v\otimes u$. Since the inner product on $\mathcal K$ is positive definite and the eigenspaces of $L_0$ are finite dimensional, $\mathcal K$ decomposes into an orthogonal sum of irreducible highest-weight representations of the total algebra: at each level $N$ one removes the descendants of the primaries found at lower levels and takes the orthogonal complement, which is annihilated by all $L_{n>0}$. This produces orthonormal primaries $u_{N,j}$ of weight $h+N$, $N=0,1,2,\dots$ Within each family, the Ward identity fixes all amplitudes in terms of $\lambda_{N,j}=\widetilde A(u_{N,j})$, and the Gram matrices are those of the representation of weight $h+N$ at central charge $c'$, so that the family contributes $\lambda_{N,j}^2\, V^{(c')}_{h+N}(\z)$ to the state sum.\footnote{For $c'>1$ these are Verma modules without null states. If $c'\leq1$, as for two decoupled Ising models, null states of the total algebra can appear. They are zero in $\mathcal K$, and $V^{(c')}_{h+N}$ is then the block of the irreducible representation.} Summing over the families,
\begin{equation}\label{eq:alpha-sum}
    \alpha_N=\sum_j\lambda_{N,j}^2\geq0\ec
\end{equation}
and $\alpha_0=1$ because $\ket h\otimes\ket0$ is the only state of weight $h$. Each $\alpha_N$ involves finitely many states, so no convergence argument is needed for the positivity. When the external weights are pairwise equal, the block is invariant, up to the phase $(-1)^h$, under the exchange of the two operators of a pair, $\z\to\z/(\z-1)$~\cite{Perlmutter:2015iya}. Each block on the right-hand side of~\eqref{eq:branching} picks up $(-1)^{h+N}$, so $\alpha_N=0$ for odd $N$. Finally, $V^{(c)}_h(\z)$ is analytic in the plane cut along $[1,\infty)$, as is manifest in the elliptic representation~\eqref{eq:chiral-prefactor}, so the expansion~\eqref{eq:state-sum} converges for $|\z|<1$. For real $0<\z<1$ all the terms of the double series obtained by inserting~\eqref{eq:state-sum} in~\eqref{eq:branching} are non-negative, so the two sums can be exchanged, and the series~\eqref{eq:branching} converges absolutely and uniformly on compact subsets of the unit disk. It can therefore be differentiated term by term, which is all the bootstrap at $\z=\zb=1/2$ requires. 

\paragraph{The first new primaries.} The construction has a transparent meaning when both theories exist. The operator $\cO\otimes\id$ is a primary of the total stress tensor, but its descendants under the two separate algebras contain further primaries of the total one, the $u_{N,j}$, and $\alpha_N$ is the sum of their squared OPE coefficients, $\lambda^2_{\sigma\sigma u_{N,j}}/\lambda^2_{\sigma\sigma\cO}$. The simplest example is in the vacuum sector. Besides the total stress tensor $T+T_s$, the product theory contains the orthogonal combination
\begin{equation}\label{eq:W}
    W=T-\frac{c}{s}\,T_s\ec
\end{equation}
which is annihilated by $L_1$ and $L_2$ and is therefore a Virasoro primary of weight two: a second conserved spin-two current, as expected in a decoupled theory. Its norm is $\braket{W|W}=cc'/(2s)$ and its amplitude is $\widetilde A(W)=h_\sigma$, so that
\begin{equation}\label{eq:alpha2-vac}
    \alpha_2^{\rm vac}=\frac{2h_\sigma^2\,s}{c\,c'}=\frac{\Delta_\sigma^2\,(c'-c)}{2\,c\,c'}\ed
\end{equation}
For a generic weight $h$, the same computation at level two gives
\begin{equation}\label{eq:alpha2}
    \alpha_2=\frac{(c'-c)\big[h(h+4h_\sigma-1)+2h_\sigma\big]^2}{2\,P_c(h)\,P_{c'}(h)}\ec\qquad P_c(h)=(4h-1)^2+(c-1)(2h+1)\ec
\end{equation}
with $P_c(h)>0$ for $c>1$ and $h\geq0$. Note that $\alpha_2$ is negative for $c'<c$: the expansion can never be positive in that direction. The higher coefficients follow from~\eqref{eq:branching} by comparing power series: if $V^{(c)}_h(\z)=\z^h\sum_n f_n(h,c)\z^n$, then $f_n(h,c)=\sum_{N\leq n}\alpha_N f_{n-N}(h+N,c')$. The coefficients $\alpha_{N}$ can be found by inverting this linear triangular system, using $f_0=1$.

\paragraph{Remarks.} The argument uses only three ingredients: a positive inner product on the exchanged representation, the Ward identity~\eqref{eq:ward}, and a change of orthonormal basis. For two pairs with different weights, as in $\braket{\phi_1\phi_1\phi_2\phi_2}$, the same construction gives an identity of the form~\eqref{eq:branching}, but with coefficients $\sum_j\widetilde A_{12}(u_{N,j})\widetilde A_{34}(u_{N,j})$ that have no definite sign; already $\alpha_2$ changes sign. In a mixed-correlator system, however, these coefficients form positive semi-definite matrices in the space of OPE coefficients, since each primary of the total algebra contributes a rank-one term, and the conclusions below carry over to the semi-definite positivity conditions of such a system. Next, as $s\to\infty$ the blocks at $c'$ reduce to global blocks, and~\eqref{eq:branching} becomes the expansion of a Virasoro block in global blocks $g_{h+N}(\z)=\z^{h+N}\hypgeo{2}{1}(h+N,h+N;2h+2N;\z)$ with non-negative coefficients. This also follows directly from~\eqref{eq:state-sum} by summing over the quasi-primaries of the representation, $L_1u=0$, instead of the primaries. Finally, the spectator operator need not be the identity: tensoring with a primary of weight $h'$ in a spectator theory with external weight $h'_\sigma$, one finds that the products $V^{(c)}_hV^{(s)}_{h'}$ expand with non-negative coefficients in blocks $V^{(c+s)}_{h+h'+N}$ with external weight $h_\sigma+h'_\sigma$. This is how the product theories of Table~\ref{tab:known-cfts} fit in the bounds.

\subsection{Nested allowed regions}\label{sec:nested}

For the full blocks~\eqref{eq:virasoro-block-full}, applying~\eqref{eq:branching} to both chiral factors gives
\begin{equation}\label{eq:branching-2d}
    \cG^{(c)}_{\Delta,\ell}(\z,\zb)=\sum_{N,M\ {\rm even}}\alpha_N(h)\,\alpha_M(\bar h)\,\frac{(-2)^{\ell'}(1+\delta_{\ell',0})}{(-2)^{\ell}(1+\delta_{\ell,0})}\;\cG^{(c')}_{\Delta+N+M,\,\ell'}(\z,\zb)\ec\qquad \ell'=\abs{\ell+N-M}\ec
\end{equation}
with $h=(\Delta+\ell)/2$ and $\bar h=(\Delta-\ell)/2$. All the coefficients are non-negative, because $\ell'-\ell$ is even. Consider then a solution of the crossing equation~\eqref{eq:sum-rule} at central charge $c$, with a spectrum of Virasoro primaries $\{(\Delta,\ell)\}$ in $\sigma\times\sigma$ and squared OPE coefficients $\lambda^2_{\sigma\sigma\cO}\geq0$. Inserting~\eqref{eq:branching-2d} gives a solution at $c'$ with the same $\Delta_\sigma$, non-negative squared OPE coefficients, and the spectrum
\begin{equation}\label{eq:images}
    (\Delta,\ell)\ \longrightarrow\ \big(\Delta+N+M,\;\abs{\ell+N-M}\big)\ec\qquad N,M=0,2,4,\dots
\end{equation}
If both theories exist, this is just the OPE $\sigma\times\sigma$ of the product theory, organised in representations of the total Virasoro algebra. In particular, the identity generates the operators $(h,\bar h)=(2,0)$, $(0,2)$ and $(2,2)$, with coefficients $\alpha_2^{\rm vac}$ and $(\alpha_2^{\rm vac})^2$: the current $W$, its antiholomorphic partner, and the scalar $W\bar W$ of dimension four.

We now compare with the assumptions of Section~\ref{sec:cft-desert}: all scalars in $\sigma\times\sigma$ other than the identity have $\Delta\geq\Delta_\eps$, and spinning operators have $\Delta\geq\ell$. The images~\eqref{eq:images} have even spin, satisfy unitarity, and are never lighter than their parent. A scalar descends from a parent of spin $\ell$ only if $M-N=\ell$, in which case $\Delta'\geq\Delta+\ell\geq2\ell$; since $\ell=0$ or $\ell\geq2$, the new scalars have either $\Delta'\geq\Delta_\eps$ or $\Delta'\geq4$, and the vacuum contributes scalars with $\Delta'=4,6,\dots$ Therefore
\begin{equation}\label{eq:nested}
    (\Delta_\sigma,\Delta_\eps)\ \text{allowed at}\ c\ \text{ and }\ \Delta_\eps\leq4\qquad\Longrightarrow\qquad(\Delta_\sigma,\Delta_\eps)\ \text{allowed at}\ c'=c+s\ec
\end{equation}
for every $s$ in~\eqref{eq:spectator-charges}, and the bounds of Figure~\ref{fig:z2} satisfy
\begin{equation}\label{eq:bound-monotonic}
    \Delta_\eps^{\max}(c+s,\Delta_\sigma)\geq\min\big(\Delta_\eps^{\max}(c,\Delta_\sigma),\,4\big)\ed
\end{equation}
This holds not only for actual CFTs but at the level of the numerical bounds, at any derivative order. If a functional $\omega$ excludes $(\Delta_\sigma,\Delta_\eps)$ at $c'$, that is $\omega(\cF^{(c')}_{0,0})>0$ and $\omega(\cF^{(c')}_{\Delta,\ell})\geq0$ for all operators allowed at $c'$, then by~\eqref{eq:branching-2d}, applied to the crossing functions~\eqref{eq:crossing-function}, the same functional satisfies $\omega(\cF^{(c)}_{0,0})\geq\omega(\cF^{(c')}_{0,0})>0$ and $\omega(\cF^{(c)}_{\Delta,\ell})\geq0$ for all operators allowed at $c$, since these are non-negative combinations of operators allowed at $c'$. The same $\omega$ excludes the point at $c$. Strictly speaking, $\omega$ must be non-negative on all even spins, since the images~\eqref{eq:images} of a given operator have arbitrarily large spin, while positivity is imposed numerically on a finite set of spins (Table~\ref{tab:z2-parameters}); as usual, we assume that the functionals found remain positive at larger spins.

\subsection{Comparison with the numerical bounds}\label{sec:comparison}

The bounds of Figures~\ref{fig:z2-c1p001} and~\ref{fig:z2} can be compared pair by pair. For each pair we state what the branching rule guarantees and what the data show. The rule applies to the following pairs.
\begin{itemize}
\item \emph{$c'=12/5$ and the unitary theories with $c\leq1$} ($s\geq7/5$). Every solution of crossing realised at $c\leq1$ must be allowed at $c=12/5$: the Ising point $(1/8,1)$, the minimal models $\mathcal M(m+1,m)$ with $\sigma=\phi_{1,2}$, at $(\Delta_\sigma,\Delta_\eps)=\big(\tfrac{m-2}{2(m+1)},\tfrac{2(m-1)}{m+1}\big)$, and the compact-boson line $\Delta_\eps=\min(4\Delta_\sigma,2)$. The data show more than inclusion. The Ising point is the kink of the bound, the minimal-model points lie on the boundary within the resolution of the scan, and so does their accumulation point $(1/2,2)$, where the bound touches the compact-boson line. This saturation is not a consequence of the branching rule.
\item \emph{$c'=12/5$ and $c=1.001$, $1.01$, $8/7$, $7/5$} ($s=1399/1000$, $139/100$, $44/35$, $1$). The rule requires the bound at $c=12/5$ to lie above the other four wherever these are below four, and it does. Above four the rule says nothing, and indeed for $\Delta_\sigma\gtrsim1$ the regions allowed at $c=8/7$ and $7/5$ extend to $\Delta_\eps\simeq5$, where the bound at $c=12/5$ lies between four and five. This is compatible with~\eqref{eq:bound-monotonic} only because of the threshold at $\Delta_\eps=4$, that is, because of the scalar $W\bar W$.
\item \emph{$c'=7/5$ and the tricritical Ising model} ($s=7/10$). The point $(1/5,6/5)$, which is the minimal model $\mathcal M(5,4)$ of the family above, must be allowed at $c=7/5$. It lies on the boundary.
\end{itemize}
For the following pairs no inclusion is guaranteed.
\begin{itemize}
\item \emph{$c=1.001$, $1.01$ and the compact boson at $c=1$} ($s=10^{-3}$, $10^{-2}$). The compact-boson line is excluded wherever the bound collapses, for $0.17\lesssim\Delta_\sigma\lesssim0.96$ at $c=1.001$. At $c=1.01$, the bound follows the compact-boson line up to $\Delta_\sigma\approx0.16$, then drops below it before returning to $\Delta_\eps=2$ at $\Delta_\sigma\approx0.97$. Had the branching rule held for every $s>0$, the bound at $c=1+\epsilon$ would have stayed above $\min(4\Delta_\sigma,2)$ for every $\epsilon$, and the desert could not exist. The collapse is possible precisely because there is no unitary vacuum representation of the Virasoro algebra, let alone a unitary CFT, with central charge $\epsilon$.
\item \emph{$c=1.01$, $8/7$ and the minimal models with $m\geq4$} ($s<1/2$). These models are excluded at both central charges. The Ising model, with $s=0.51$ and $9/14$, is not covered by the rule either, and survives at the top of the spike (Section~\ref{app:at-point}).
\item \emph{$c'=7/5$ and the minimal models other than $\mathcal M(5,4)$} ($s=7/5-c_m$ is not an admissible increment). The models with $5\leq m\leq7$ lie on or inside the allowed region within the resolution of the scan, whereas those with $m\geq8$ lie above the bound. The Ising model, with $s=9/10$, again sits at the top of the spike.
\item \emph{$c'=14/5$ and $c=12/5$} ($s=2/5$). We also computed the bound at $c=14/5$ (not shown) and observed no change from $c=12/5$ within our numerical resolution, although no inclusion is guaranteed between them. The reason is that the first branching coefficients that can be negative are very small, as we now explain.
\end{itemize}

The failure of positivity for $s$ outside~\eqref{eq:spectator-charges} can be seen directly in the coefficients. For $0<s<1/2$ the vacuum representation at central charge $s$ has a negative-norm state at level six, the continuation of the Ising null vector~\cite{Friedan:1983xq}, so the construction of Section~\ref{sec:branching-proof} breaks down at level six, and indeed $\alpha_6$ can be negative: for the vacuum block with $c=3/2$, $s=1/100$ and $\Delta_\sigma=1/5$, an exact computation from the Gram matrices gives $\alpha_2>0$, $\alpha_4>0$ and
\begin{equation}
    \alpha_6=-\frac{1060321750443}{4502914241536351468750}\simeq-2.4\times10^{-10}\ed
\end{equation}
More generally, for $c_m<s<c_{m+1}$ the vacuum representation at central charge $s$ is positive definite below level $m(m+1)$~\cite{Friedan:1983xq}, the argument goes through up to that level, and $\alpha_N\geq0$ for $N<m(m+1)$. The first coefficients that can be negative are therefore very small, which explains why the expansion can appear positive in numerical tests at low orders, and why bounds with no guaranteed inclusion can nevertheless coincide within the resolution of the scan.

Finally, every image~\eqref{eq:images} of a solution at lower central charge contains the spin-two primary $W$ at $\Delta=2$, with the coefficient~\eqref{eq:alpha2-vac}. Imposing a twist gap in the spin-two sector, $\Delta>2$ at $\ell=2$, removes all these solutions from the allowed region, and the bounds no longer need to be nested in $c$. The dependence of the spike at $(1/8,1)$ on such a gap would be a sharp test of its origin.

\subsection{The limit $c\to1^+$ and the spike at $\Delta_\sigma=1/8$}\label{app:at-point}

The spectator argument says nothing about increments below $1/2$, and in particular about the relation between the bound at $c=1.01$ and the theories at $c=1$. Two exact properties of the blocks clarify this limit. The first concerns the singular behaviour of the blocks as $c\to1^+$ at degenerate exchanged weights. For $1<c<25$ the off-diagonal Kac weights $h_{m,n}(c)$ of~\eqref{eq:H-recursion} are complex, and as $c\to1^+$ they condense in conjugate pairs onto the quarter-squares $h_{m,n}(1)=(m-n)^2/4$, where the Kac determinant acquires double zeros. The blocks do not remain finite there. At level two, with $P_c(h)$ as in~\eqref{eq:alpha2}, one finds for $V^{(c)}_h(\z)=\z^h\big(1+\tfrac h2\z+f_2(h)\z^2+\dots\big)$
\begin{equation}\label{eq:f2-peak}
    f_2\big(\tfrac14\big)=\frac{2\,(h_\sigma-\tfrac1{16})^2}{c-1}+\frac{25}{384}\ec
\end{equation}
so that for generic $h_\sigma$ the block at $h=1/4$ grows like $1/(c-1)$; the same happens at $h=9/4$, $25/4,\dots$ at higher levels. The exception is $h_\sigma=1/16$, that is $\Delta_\sigma=1/8$, where the numerators vanish. At $c=1$ and $h_\sigma=1/16$, for non-degenerate exchanged weight $h$, the Virasoro blocks are the Ashkin--Teller functions of Appendix~\ref{app:known-blocks}, $V_h(\z)=(16q)^h\Phi(\z)$ with $\Phi(\z)=(1-\z)^{-1/8}\vartheta_3(q)^{-1}$. These functions extend to entire functions of $h$; at degenerate weights, the continuation gives the $U(1)$ blocks~\cite[Section~5.2]{Nemkov:2021huu}. In the language of~\eqref{eq:chiral-prefactor}, $H\equiv1$ for this entire continuation. The external dimension $\Delta_\sigma=1/8$ is therefore the only one at which the blocks, and with them the bootstrap problem, are continuous at $c=1$.

The second property concerns which four-point functions of $c=1$ theories admit a positive expansion in blocks at $c=1+\epsilon$. We computed these expansions exactly, from the Gram matrices of the Verma modules, order by order in $\z$. For the vertex operators of the compact boson at generic radius, the four-point function is elementary: up to the prefactor, it is a sum of products of $(1-\z)^{\pm2h_\sigma}$ and of $\z^{4h_\sigma}(1-\z)^{-2h_\sigma}$, the latter being the $c=1$ block of the lightest charged state, of weight $4h_\sigma$. Expanded in blocks at $c=1.01$, these functions have negative coefficients, at level six for $\Delta_\sigma=1/4$, $1/2$ and $3/4$ and at level ten for $\Delta_\sigma=1$, so nothing protects the compact-boson line, and the bound is free to drop below it, as it does. At $\Delta_\sigma=1/8$ the relevant theories are those of the orbifold branch, whose twist-field correlators are lattice sums, with positive coefficients, of the products $(16q)^h\Phi(\z)\,(16\bar q)^{\bar h}\Phi(\zb)$ (Appendix~\ref{app:orbifold}). At degenerate exchanged weights, $(16q)^h\Phi$ is the $U(1)$ block rather than an individual irreducible Virasoro block: $\Phi-V^{(1)}_0$ starts at order $\z^4$ with a positive coefficient, the contribution of the first even discrete state, and at the degenerate weights $h=1/4,1,9/4,\dots$ the $U(1)$ block differs from the block of the irreducible Virasoro representation. We find that every $U(1)$ block we tested, $h=0,\tfrac14,\tfrac3{10},\tfrac7{10},1,\tfrac{13}{10},\tfrac94,3,4$, expands with positive coefficients in blocks at $c=1.01$ through level twelve ($\Phi$ also at $c=1.001$ and $1.1$), whereas the irreducible Virasoro blocks at $h=1/4$, $1$ and $9/4$ alone do not. Since the orbifold correlators contain these weights only through $U(1)$ modules, the whole orbifold branch lifts to $c=1+\epsilon$. For two decoupled Ising models, which sit on this branch, we checked the Ising blocks directly, $V_0=(1-\z)^{-1/8}\big[(1+\sqrt{1-\z})/2\big]^{1/2}$ and $V_{1/2}=2(1-\z)^{-1/8}\big[(1-\sqrt{1-\z})/2\big]^{1/2}$: all coefficients are positive through level sixteen at $c=1.01$, through level fourteen at $c=1.001$, and through level twelve at $c=1.05$, $1.1$, $8/7$, $6/5$, $7/5$ and $3/2$. For $c=6/5$ and $3/2$, the increment $s=c-\tfrac12$ is a unitary central charge, and positivity holds to all orders by the argument of the previous sections. For the other values, $s\in(\tfrac12,\tfrac7{10})$ or $s=\tfrac9{10}\in(c_7,c_8)$, the same argument guarantees the first eleven, respectively fifty-five, levels, since the vacuum representation at central charge $s$ is positive definite below level $m(m+1)$; the remaining ones are checked, not proven.

The bound at $\Delta_\sigma=1/8$ can therefore not fall below the largest gap realised on the orbifold branch. Along the branch, the lightest scalar in $\sigma\times\sigma$ is the lightest of the momentum and winding states appearing in the OPE, since the even discrete states have $\Delta\geq8$; its dimension is largest, and equal to one, at the radius where the two are degenerate, which is the point of two decoupled Ising models. The spike thus reaches exactly $\Delta_\eps=1$, as observed at $c=1.001$, $1.01$, $8/7$ and $7/5$, and it is narrow because no other $\Delta_\sigma$ enjoys the same continuity. What this analysis does not explain is the collapse of the bound between the spike and $\Delta_\sigma\simeq1$, nor its return to the value $\Delta_\eps=2$ of the marginal operator beyond that point. We only note that in the collapsed window the bound is of order $c-1$: about $0.004$ at $c=1.01$ and about $0.08$ at $c=8/7$.

\section{Spectrum of the decoupled theories}\label{app:spectrum}
{\footnotesize\renewcommand{\arraystretch}{1.25}
\begin{longtable}{c c p{12.2cm}}
\caption{Virasoro primaries for the theory of three decoupled Ising models ($c=\tfrac32$, $\Delta\leq10$), in the sectors of $S_3\ltimes(\Z_2)^3$ exchanged in $\sigma_i\times\sigma_j$ ($S$, $X$, $Z$, $B$), by sector and spin $\ell=|h-\bar h|$. $\Delta^{(N)}$ denotes the dimension and multiplicity, if $N>1$. The irreps appearing in the $\sigma_i\times\sigma_j$ OPE are denoted in bold. The naming conventions followed for irreps are those of \cite{Kousvos:2021rar} which deals with generic replica theories $S_n \ltimes G^n$ ($G$ is any group).}\label{tab:ising-spectrum}\\
\toprule sector & $\ell$ & $\Delta^{(N)}$ \\ \midrule \endfirsthead
\toprule sector & $\ell$ & $\Delta^{(N)}$ \\ \midrule \endhead
\bottomrule \endlastfoot
$S$ & 0 & $\mathbf{0},\ \mathbf{1},\ 2,\ 3,\ \mathbf{4}^{(2)},\ 5^{(2)},\ 6,\ 7^{(2)},\ \mathbf{8}^{(3)},\ \mathbf{9}^{(6)},\ 10^{(6)}$ \\
 & 1 & $6^{(2)},\ 8,\ 9,\ 10^{(3)}$ \\
 & 2 & $4,\ \mathbf{6}^{(2)},\ 7,\ 8^{(2)},\ 9^{(2)},\ \mathbf{10}^{(4)}$ \\
 & 3 & $7,\ 8^{(2)},\ 9,\ 10^{(3)}$ \\
 & 4 & $\mathbf{4},\ \mathbf{5}^{(2)},\ 6^{(2)},\ \mathbf{8}^{(3)},\ 9^{(3)},\ 10^{(2)}$ \\
 & 5 & $6,\ 7,\ 9^{(3)},\ 10^{(3)}$ \\
 & 6 & $\mathbf{6},\ \mathbf{7}^{(2)},\ 8^{(2)},\ 9,\ \mathbf{10}^{(6)}$ \\
 & 7 & $8^{(2)},\ 9^{(2)},\ 10$ \\
 & 8 & $\mathbf{8}^{(2)},\ \mathbf{9}^{(4)},\ 10^{(4)}$ \\
 & 9 & $10^{(3)}$ \\
 & 10 & $\mathbf{10}^{(2)}$ \\
\midrule
$X$ & 0 & $\mathbf{1},\ 2,\ \mathbf{4}^{(2)},\ 5^{(2)},\ 6,\ 7^{(2)},\ \mathbf{8}^{(4)},\ \mathbf{9}^{(9)},\ 10^{(9)}$ \\
 & 1 & $3,\ 4,\ 5,\ 6^{(2)},\ 7,\ 8^{(4)},\ 9^{(4)},\ 10^{(7)}$ \\
 & 2 & $\mathbf{2},\ 3,\ 4,\ 5,\ \mathbf{6}^{(3)},\ 7^{(4)},\ 8^{(3)},\ 9^{(3)},\ \mathbf{10}^{(7)}$ \\
 & 3 & $4,\ 5,\ 7^{(4)},\ 8^{(3)},\ 9^{(3)},\ 10^{(6)}$ \\
 & 4 & $\mathbf{4},\ \mathbf{5}^{(3)},\ 6^{(3)},\ 7,\ \mathbf{8}^{(6)},\ 9^{(6)},\ 10^{(4)}$ \\
 & 5 & $6^{(2)},\ 7^{(3)},\ 8,\ 9^{(6)},\ 10^{(7)}$ \\
 & 6 & $\mathbf{6},\ \mathbf{7}^{(4)},\ 8^{(4)},\ 9,\ \mathbf{10}^{(10)}$ \\
 & 7 & $7,\ 8^{(4)},\ 9^{(5)},\ 10^{(2)}$ \\
 & 8 & $\mathbf{8}^{(3)},\ \mathbf{9}^{(7)},\ 10^{(7)}$ \\
 & 9 & $9,\ 10^{(7)}$ \\
 & 10 & $\mathbf{10}^{(4)}$ \\
\midrule
$Z$ & 0 & $\mathbf{\tfrac{1}{4}},\ \tfrac{5}{4},\ \mathbf{\tfrac{9}{4}},\ \tfrac{13}{4}^{(2)},\ \mathbf{\tfrac{17}{4}},\ \tfrac{21}{4}^{(2)},\ \mathbf{\tfrac{25}{4}}^{(5)},\ \tfrac{29}{4}^{(5)},\ \mathbf{\tfrac{33}{4}}^{(5)},\ \tfrac{37}{4}^{(8)}$ \\
 & 1 & $\tfrac{9}{4},\ \tfrac{17}{4}^{(2)},\ \tfrac{21}{4},\ \tfrac{25}{4}^{(3)},\ \tfrac{29}{4}^{(4)},\ \tfrac{33}{4}^{(6)},\ \tfrac{37}{4}^{(7)}$ \\
 & 2 & $\mathbf{\tfrac{9}{4}},\ \tfrac{13}{4},\ \mathbf{\tfrac{17}{4}}^{(2)},\ \tfrac{21}{4}^{(3)},\ \mathbf{\tfrac{25}{4}}^{(2)},\ \tfrac{29}{4}^{(4)},\ \mathbf{\tfrac{33}{4}}^{(8)},\ \tfrac{37}{4}^{(8)}$ \\
 & 3 & $\tfrac{13}{4},\ \tfrac{17}{4},\ \tfrac{21}{4},\ \tfrac{25}{4}^{(4)},\ \tfrac{29}{4}^{(2)},\ \tfrac{33}{4}^{(5)},\ \tfrac{37}{4}^{(10)}$ \\
 & 4 & $\mathbf{\tfrac{17}{4}}^{(2)},\ \tfrac{21}{4}^{(2)},\ \mathbf{\tfrac{25}{4}}^{(3)},\ \tfrac{29}{4}^{(5)},\ \mathbf{\tfrac{33}{4}}^{(4)},\ \tfrac{37}{4}^{(7)}$ \\
 & 5 & $\tfrac{21}{4}^{(2)},\ \tfrac{25}{4}^{(2)},\ \tfrac{29}{4}^{(3)},\ \tfrac{33}{4}^{(7)},\ \tfrac{37}{4}^{(3)}$ \\
 & 6 & $\mathbf{\tfrac{25}{4}}^{(4)},\ \tfrac{29}{4}^{(4)},\ \mathbf{\tfrac{33}{4}}^{(5)},\ \tfrac{37}{4}^{(10)}$ \\
 & 7 & $\tfrac{29}{4}^{(3)},\ \tfrac{33}{4}^{(5)},\ \tfrac{37}{4}^{(5)}$ \\
 & 8 & $\mathbf{\tfrac{33}{4}}^{(6)},\ \tfrac{37}{4}^{(6)}$ \\
 & 9 & $\tfrac{37}{4}^{(6)}$ \\
\midrule
$B$ & 0 & $\tfrac{13}{4}^{(2)},\ \tfrac{21}{4}^{(2)},\ \tfrac{25}{4}^{(4)},\ \tfrac{29}{4}^{(4)},\ \tfrac{33}{4}^{(4)},\ \tfrac{37}{4}^{(8)}$ \\
 & 1 & $\mathbf{\tfrac{5}{4}},\ \tfrac{9}{4},\ \mathbf{\tfrac{13}{4}},\ \tfrac{17}{4}^{(2)},\ \mathbf{\tfrac{21}{4}}^{(2)},\ \tfrac{25}{4}^{(3)},\ \mathbf{\tfrac{29}{4}}^{(5)},\ \tfrac{33}{4}^{(6)},\ \mathbf{\tfrac{37}{4}}^{(8)}$ \\
 & 2 & $\tfrac{13}{4},\ \tfrac{17}{4},\ \tfrac{21}{4}^{(3)},\ \tfrac{25}{4},\ \tfrac{29}{4}^{(4)},\ \tfrac{33}{4}^{(7)},\ \tfrac{37}{4}^{(7)}$ \\
 & 3 & $\mathbf{\tfrac{13}{4}}^{(2)},\ \tfrac{17}{4}^{(2)},\ \mathbf{\tfrac{21}{4}}^{(2)},\ \tfrac{25}{4}^{(4)},\ \mathbf{\tfrac{29}{4}}^{(3)},\ \tfrac{33}{4}^{(5)},\ \mathbf{\tfrac{37}{4}}^{(11)}$ \\
 & 4 & $\tfrac{17}{4},\ \tfrac{21}{4}^{(2)},\ \tfrac{25}{4}^{(2)},\ \tfrac{29}{4}^{(5)},\ \tfrac{33}{4}^{(3)},\ \tfrac{37}{4}^{(7)}$ \\
 & 5 & $\mathbf{\tfrac{21}{4}}^{(3)},\ \tfrac{25}{4}^{(3)},\ \mathbf{\tfrac{29}{4}}^{(4)},\ \tfrac{33}{4}^{(7)},\ \mathbf{\tfrac{37}{4}}^{(5)}$ \\
 & 6 & $\tfrac{25}{4}^{(3)},\ \tfrac{29}{4}^{(3)},\ \tfrac{33}{4}^{(3)},\ \tfrac{37}{4}^{(10)}$ \\
 & 7 & $\mathbf{\tfrac{29}{4}}^{(5)},\ \tfrac{33}{4}^{(5)},\ \mathbf{\tfrac{37}{4}}^{(6)}$ \\
 & 8 & $\tfrac{33}{4}^{(5)},\ \tfrac{37}{4}^{(6)}$ \\
 & 9 & $\mathbf{\tfrac{37}{4}}^{(8)}$ \\
\end{longtable}}

{\footnotesize\renewcommand{\arraystretch}{1.25}
\begin{longtable}{l c c c p{10.6cm}}
\caption{Virasoro primaries for the theory of three decoupled 3-state Potts models ($c=\tfrac{12}5$, $\Delta\leq6$), by irrep of $S_3\ltimes(S_3)^3$ and spin $\ell=|h-\bar h|$. $\Delta^{(N)}$ denotes the dimension and multiplicity, if $N>1$. The naming conventions for irreps are those of \cite{Kousvos:2024dlz}.}\label{tab:potts-spectrum}\\
\toprule irrep & dim & $n_V$ & $\ell$ & $\Delta^{(N)}$ \\ \midrule \endfirsthead
\toprule irrep & dim & $n_V$ & $\ell$ & $\Delta^{(N)}$ \\ \midrule \endhead
\bottomrule \endlastfoot
$S$ & 1 & 0 & 0 & $0,\ \tfrac{4}{5},\ \tfrac{8}{5},\ \tfrac{12}{5},\ \tfrac{14}{5},\ \tfrac{18}{5}^{(2)},\ 4,\ \tfrac{22}{5}^{(2)},\ \tfrac{24}{5},\ \tfrac{28}{5}^{(3)},\ 6$ \\
 &  &  & 1 & $\tfrac{23}{5},\ \tfrac{27}{5}^{(2)},\ \tfrac{29}{5}$ \\
 &  &  & 2 & $\tfrac{18}{5},\ \tfrac{24}{5},\ \tfrac{28}{5}^{(4)},\ 6$ \\
 &  &  & 3 & $\tfrac{19}{5},\ \tfrac{23}{5},\ \tfrac{27}{5}$ \\
 &  &  & 4 & $4,\ \tfrac{24}{5}^{(2)},\ \tfrac{28}{5}^{(3)}$ \\
 &  &  & 5 & $\tfrac{29}{5}^{(2)}$ \\
 &  &  & 6 & $6^{(2)}$ \\
\midrule
$X$ & 2 & 0 & 0 & $\tfrac{4}{5},\ \tfrac{8}{5},\ \tfrac{14}{5},\ \tfrac{18}{5}^{(3)},\ 4,\ \tfrac{22}{5}^{(2)},\ \tfrac{24}{5},\ \tfrac{28}{5}^{(4)},\ 6$ \\
 &  &  & 1 & $\tfrac{13}{5},\ \tfrac{17}{5},\ \tfrac{23}{5}^{(3)},\ \tfrac{27}{5}^{(3)},\ \tfrac{29}{5}^{(2)}$ \\
 &  &  & 2 & $2,\ \tfrac{14}{5},\ \tfrac{18}{5},\ \tfrac{22}{5},\ \tfrac{24}{5}^{(2)},\ \tfrac{28}{5}^{(7)},\ 6^{(2)}$ \\
 &  &  & 3 & $\tfrac{19}{5}^{(2)},\ \tfrac{23}{5}^{(3)},\ \tfrac{27}{5},\ \tfrac{29}{5}$ \\
 &  &  & 4 & $4,\ \tfrac{24}{5}^{(3)},\ \tfrac{28}{5}^{(5)}$ \\
 &  &  & 5 & $\tfrac{29}{5}^{(4)}$ \\
 &  &  & 6 & $6^{(2)}$ \\
\midrule
$\overline{XX}$ & 1 & 0 & 0 & $\tfrac{18}{5},\ 4,\ \tfrac{22}{5},\ \tfrac{28}{5}$ \\
 &  &  & 1 & $\tfrac{13}{5},\ \tfrac{23}{5}^{(2)},\ \tfrac{27}{5}^{(2)},\ \tfrac{29}{5}$ \\
 &  &  & 2 & $\tfrac{14}{5},\ \tfrac{24}{5},\ \tfrac{28}{5}^{(3)},\ 6$ \\
 &  &  & 3 & $\tfrac{19}{5},\ \tfrac{23}{5}^{(2)},\ \tfrac{27}{5},\ \tfrac{29}{5}$ \\
 &  &  & 4 & $\tfrac{24}{5},\ \tfrac{28}{5}^{(2)}$ \\
 &  &  & 5 & $5,\ \tfrac{29}{5}^{(2)}$ \\
 &  &  & 6 & $6$ \\
\midrule
$A$ & 3 & 0 & 0 & $\tfrac{18}{5}^{(2)},\ \tfrac{22}{5}^{(2)},\ \tfrac{28}{5}^{(4)}$ \\
 &  &  & 1 & $\tfrac{9}{5},\ \tfrac{13}{5},\ \tfrac{17}{5},\ \tfrac{23}{5}^{(3)},\ 5,\ \tfrac{27}{5}^{(4)},\ \tfrac{29}{5}^{(2)}$ \\
 &  &  & 2 & $\tfrac{18}{5},\ \tfrac{22}{5},\ \tfrac{24}{5},\ \tfrac{28}{5}^{(6)}$ \\
 &  &  & 3 & $3,\ \tfrac{19}{5}^{(2)},\ \tfrac{23}{5}^{(3)},\ \tfrac{27}{5}^{(2)},\ \tfrac{29}{5}^{(3)}$ \\
 &  &  & 4 & $\tfrac{24}{5},\ \tfrac{28}{5}^{(4)}$ \\
 &  &  & 5 & $5,\ \tfrac{29}{5}^{(5)}$ \\
 &  &  & 6 & $6$ \\
\midrule
$XA$ & 3 & 0 & 0 & $\tfrac{18}{5}^{(2)},\ \tfrac{22}{5}^{(2)},\ \tfrac{28}{5}^{(4)}$ \\
 &  &  & 1 & $\tfrac{13}{5},\ \tfrac{19}{5},\ \tfrac{23}{5}^{(3)},\ 5,\ \tfrac{27}{5}^{(4)},\ \tfrac{29}{5}^{(2)}$ \\
 &  &  & 2 & $\tfrac{18}{5},\ \tfrac{22}{5},\ \tfrac{24}{5},\ \tfrac{28}{5}^{(7)}$ \\
 &  &  & 3 & $\tfrac{19}{5}^{(2)},\ \tfrac{23}{5}^{(2)},\ \tfrac{27}{5},\ \tfrac{29}{5}^{(2)}$ \\
 &  &  & 4 & $\tfrac{24}{5}^{(2)},\ \tfrac{28}{5}^{(4)}$ \\
 &  &  & 5 & $5,\ \tfrac{29}{5}^{(4)}$ \\
 &  &  & 6 & $6$ \\
\midrule
$AA$ & 3 & 0 & 0 & $\tfrac{18}{5},\ \tfrac{22}{5},\ \tfrac{28}{5}^{(3)},\ 6$ \\
 &  &  & 1 & $\tfrac{23}{5},\ \tfrac{27}{5}^{(3)}$ \\
 &  &  & 2 & $\tfrac{18}{5},\ \tfrac{22}{5},\ \tfrac{24}{5},\ \tfrac{28}{5}^{(4)}$ \\
 &  &  & 3 & $\tfrac{27}{5},\ \tfrac{29}{5}$ \\
 &  &  & 4 & $\tfrac{24}{5},\ \tfrac{28}{5}^{(3)}$ \\
 &  &  & 5 & $\tfrac{29}{5}$ \\
 &  &  & 6 & $6$ \\
\midrule
$\overline{AA}$ & 3 & 0 & 0 & $\tfrac{18}{5},\ \tfrac{22}{5},\ \tfrac{28}{5},\ 6$ \\
 &  &  & 1 & $\tfrac{23}{5}^{(2)},\ \tfrac{27}{5}^{(3)}$ \\
 &  &  & 2 & $\tfrac{24}{5},\ \tfrac{28}{5}^{(3)}$ \\
 &  &  & 3 & $\tfrac{23}{5},\ \tfrac{27}{5},\ \tfrac{29}{5}$ \\
 &  &  & 4 & $\tfrac{24}{5},\ \tfrac{28}{5}^{(2)}$ \\
 &  &  & 5 & $\tfrac{29}{5}$ \\
\midrule
$AAA$ & 1 & 0 & 1 & $\tfrac{27}{5}$ \\
 &  &  & 3 & $\tfrac{27}{5}$ \\
\midrule
$\overline{AA}A$ & 2 & 0 & 1 & $\tfrac{27}{5}$ \\
\midrule
$\overline{AAA}$ & 1 & 0 & -- & $\text{none with }\Delta\leq6$ \\
\midrule
$V$ & 6 & 1 & 0 & $\tfrac{2}{15},\ \tfrac{14}{15},\ \tfrac{4}{3},\ \tfrac{26}{15},\ \tfrac{32}{15},\ \tfrac{44}{15}^{(3)},\ \tfrac{56}{15}^{(3)},\ \tfrac{62}{15}^{(4)},\ \tfrac{74}{15}^{(8)},\ \tfrac{16}{3}^{(2)},\ \tfrac{86}{15}^{(10)}$ \\
 &  &  & 1 & $\tfrac{29}{15},\ \tfrac{41}{15},\ \tfrac{47}{15},\ \tfrac{59}{15}^{(4)},\ \tfrac{71}{15}^{(5)},\ \tfrac{77}{15}^{(5)},\ \tfrac{89}{15}^{(16)}$ \\
 &  &  & 2 & $\tfrac{32}{15},\ \tfrac{44}{15}^{(2)},\ \tfrac{10}{3},\ \tfrac{56}{15}^{(2)},\ \tfrac{62}{15}^{(2)},\ \tfrac{74}{15}^{(9)},\ \tfrac{86}{15}^{(10)}$ \\
 &  &  & 3 & $\tfrac{47}{15},\ \tfrac{59}{15}^{(4)},\ \tfrac{71}{15}^{(3)},\ \tfrac{77}{15}^{(3)},\ \tfrac{89}{15}^{(13)}$ \\
 &  &  & 4 & $\tfrac{62}{15}^{(3)},\ \tfrac{74}{15}^{(7)},\ \tfrac{16}{3}^{(3)},\ \tfrac{86}{15}^{(6)}$ \\
 &  &  & 5 & $\tfrac{77}{15}^{(3)},\ \tfrac{89}{15}^{(11)}$ \\
\midrule
$XV$ & 6 & 1 & 0 & $\tfrac{14}{15},\ \tfrac{32}{15},\ \tfrac{44}{15}^{(2)},\ \tfrac{56}{15}^{(3)},\ \tfrac{62}{15}^{(4)},\ \tfrac{74}{15}^{(8)},\ \tfrac{16}{3}^{(2)},\ \tfrac{86}{15}^{(8)}$ \\
 &  &  & 1 & $\tfrac{29}{15},\ \tfrac{41}{15},\ \tfrac{47}{15},\ \tfrac{59}{15}^{(4)},\ \tfrac{71}{15}^{(5)},\ \tfrac{77}{15}^{(5)},\ \tfrac{89}{15}^{(16)}$ \\
 &  &  & 2 & $\tfrac{32}{15},\ \tfrac{44}{15}^{(2)},\ \tfrac{10}{3},\ \tfrac{56}{15},\ \tfrac{62}{15}^{(2)},\ \tfrac{74}{15}^{(8)},\ \tfrac{86}{15}^{(10)}$ \\
 &  &  & 3 & $\tfrac{47}{15},\ \tfrac{59}{15}^{(4)},\ \tfrac{13}{3},\ \tfrac{71}{15}^{(3)},\ \tfrac{77}{15}^{(3)},\ \tfrac{89}{15}^{(14)}$ \\
 &  &  & 4 & $\tfrac{62}{15}^{(2)},\ \tfrac{74}{15}^{(7)},\ \tfrac{16}{3}^{(2)},\ \tfrac{86}{15}^{(5)}$ \\
 &  &  & 5 & $\tfrac{77}{15}^{(3)},\ \tfrac{89}{15}^{(11)}$ \\
\midrule
$VA$ & 12 & 1 & 0 & $\tfrac{44}{15}^{(2)},\ \tfrac{56}{15}^{(4)},\ \tfrac{62}{15}^{(2)},\ \tfrac{74}{15}^{(8)},\ \tfrac{86}{15}^{(16)}$ \\
 &  &  & 1 & $\tfrac{29}{15},\ \tfrac{41}{15},\ \tfrac{47}{15},\ \tfrac{59}{15}^{(4)},\ \tfrac{71}{15}^{(8)},\ \tfrac{77}{15}^{(5)},\ \tfrac{89}{15}^{(20)}$ \\
 &  &  & 2 & $\tfrac{44}{15},\ \tfrac{56}{15}^{(2)},\ \tfrac{62}{15},\ \tfrac{74}{15}^{(10)},\ \tfrac{86}{15}^{(16)}$ \\
 &  &  & 3 & $\tfrac{47}{15},\ \tfrac{59}{15}^{(4)},\ \tfrac{13}{3},\ \tfrac{71}{15}^{(4)},\ \tfrac{77}{15}^{(4)},\ \tfrac{89}{15}^{(18)}$ \\
 &  &  & 4 & $\tfrac{62}{15},\ \tfrac{74}{15}^{(6)},\ \tfrac{16}{3},\ \tfrac{86}{15}^{(7)}$ \\
 &  &  & 5 & $\tfrac{77}{15}^{(3)},\ \tfrac{89}{15}^{(13)}$ \\
\midrule
$VAA$ & 6 & 1 & 0 & $\tfrac{56}{15},\ \tfrac{74}{15},\ \tfrac{86}{15}^{(8)}$ \\
 &  &  & 1 & $\tfrac{71}{15}^{(3)},\ \tfrac{89}{15}^{(4)}$ \\
 &  &  & 2 & $\tfrac{56}{15},\ \tfrac{74}{15}^{(2)},\ \tfrac{86}{15}^{(6)}$ \\
 &  &  & 3 & $\tfrac{71}{15},\ \tfrac{89}{15}^{(4)}$ \\
 &  &  & 4 & $\tfrac{74}{15},\ \tfrac{86}{15}^{(3)}$ \\
 &  &  & 5 & $\tfrac{89}{15}^{(2)}$ \\
\midrule
$V\overline{AA}$ & 6 & 1 & 0 & $\tfrac{56}{15},\ \tfrac{74}{15},\ \tfrac{86}{15}^{(6)}$ \\
 &  &  & 1 & $\tfrac{71}{15}^{(3)},\ \tfrac{89}{15}^{(4)}$ \\
 &  &  & 2 & $\tfrac{74}{15},\ \tfrac{86}{15}^{(6)}$ \\
 &  &  & 3 & $\tfrac{71}{15},\ \tfrac{89}{15}^{(4)}$ \\
 &  &  & 4 & $\tfrac{74}{15},\ \tfrac{86}{15}^{(2)}$ \\
 &  &  & 5 & $\tfrac{89}{15}^{(2)}$ \\
\midrule
$Z$ & 12 & 2 & 0 & $\tfrac{4}{15},\ \tfrac{16}{15},\ \tfrac{22}{15},\ \tfrac{34}{15}^{(2)},\ \tfrac{8}{3},\ \tfrac{46}{15}^{(3)},\ \tfrac{52}{15}^{(2)},\ \tfrac{64}{15}^{(10)},\ \tfrac{14}{3},\ \tfrac{76}{15}^{(10)},\ \tfrac{82}{15}^{(12)}$ \\
 &  &  & 1 & $\tfrac{31}{15},\ \tfrac{37}{15},\ \tfrac{49}{15}^{(3)},\ \tfrac{61}{15}^{(5)},\ \tfrac{67}{15}^{(4)},\ \tfrac{79}{15}^{(17)},\ \tfrac{17}{3}$ \\
 &  &  & 2 & $\tfrac{34}{15}^{(2)},\ \tfrac{46}{15}^{(2)},\ \tfrac{52}{15}^{(3)},\ \tfrac{64}{15}^{(7)},\ \tfrac{14}{3}^{(2)},\ \tfrac{76}{15}^{(11)},\ \tfrac{82}{15}^{(6)}$ \\
 &  &  & 3 & $\tfrac{49}{15}^{(2)},\ \tfrac{61}{15}^{(4)},\ \tfrac{67}{15}^{(4)},\ \tfrac{79}{15}^{(11)},\ \tfrac{17}{3}$ \\
 &  &  & 4 & $\tfrac{64}{15}^{(6)},\ \tfrac{76}{15}^{(8)},\ \tfrac{82}{15}^{(9)}$ \\
 &  &  & 5 & $\tfrac{79}{15}^{(7)}$ \\
\midrule
$B$ & 12 & 2 & 0 & $\tfrac{22}{15},\ \tfrac{34}{15},\ \tfrac{46}{15}^{(2)},\ \tfrac{52}{15},\ \tfrac{64}{15}^{(9)},\ \tfrac{76}{15}^{(10)},\ \tfrac{82}{15}^{(11)}$ \\
 &  &  & 1 & $\tfrac{19}{15},\ \tfrac{31}{15},\ \tfrac{37}{15},\ \tfrac{49}{15}^{(4)},\ \tfrac{11}{3},\ \tfrac{61}{15}^{(5)},\ \tfrac{67}{15}^{(4)},\ \tfrac{79}{15}^{(18)},\ \tfrac{17}{3}^{(2)}$ \\
 &  &  & 2 & $\tfrac{34}{15},\ \tfrac{46}{15}^{(2)},\ \tfrac{52}{15}^{(3)},\ \tfrac{64}{15}^{(6)},\ \tfrac{14}{3},\ \tfrac{76}{15}^{(10)},\ \tfrac{82}{15}^{(6)}$ \\
 &  &  & 3 & $\tfrac{49}{15}^{(3)},\ \tfrac{61}{15}^{(4)},\ \tfrac{67}{15}^{(4)},\ \tfrac{79}{15}^{(13)},\ \tfrac{17}{3}^{(2)}$ \\
 &  &  & 4 & $\tfrac{64}{15}^{(4)},\ \tfrac{76}{15}^{(7)},\ \tfrac{82}{15}^{(9)}$ \\
 &  &  & 5 & $\tfrac{79}{15}^{(9)}$ \\
\midrule
$AZ$ & 12 & 2 & 0 & $\tfrac{46}{15}^{(2)},\ \tfrac{64}{15}^{(4)},\ \tfrac{76}{15}^{(8)},\ \tfrac{82}{15}^{(2)}$ \\
 &  &  & 1 & $\tfrac{31}{15},\ \tfrac{49}{15},\ \tfrac{61}{15}^{(4)},\ \tfrac{67}{15},\ \tfrac{79}{15}^{(10)}$ \\
 &  &  & 2 & $\tfrac{46}{15},\ \tfrac{64}{15}^{(2)},\ \tfrac{76}{15}^{(9)},\ \tfrac{82}{15}^{(2)}$ \\
 &  &  & 3 & $\tfrac{49}{15},\ \tfrac{61}{15}^{(3)},\ \tfrac{67}{15},\ \tfrac{79}{15}^{(6)},\ \tfrac{17}{3}$ \\
 &  &  & 4 & $\tfrac{64}{15},\ \tfrac{76}{15}^{(4)},\ \tfrac{82}{15}^{(2)}$ \\
 &  &  & 5 & $\tfrac{79}{15}^{(3)}$ \\
\midrule
$AB$ & 12 & 2 & 0 & $\tfrac{46}{15}^{(2)},\ \tfrac{64}{15}^{(4)},\ \tfrac{76}{15}^{(8)},\ \tfrac{82}{15}^{(2)}$ \\
 &  &  & 1 & $\tfrac{49}{15},\ \tfrac{61}{15}^{(4)},\ \tfrac{79}{15}^{(9)}$ \\
 &  &  & 2 & $\tfrac{46}{15},\ \tfrac{64}{15}^{(3)},\ \tfrac{76}{15}^{(9)},\ \tfrac{82}{15}^{(2)}$ \\
 &  &  & 3 & $\tfrac{61}{15}^{(2)},\ \tfrac{67}{15},\ \tfrac{79}{15}^{(6)}$ \\
 &  &  & 4 & $\tfrac{64}{15},\ \tfrac{76}{15}^{(5)},\ \tfrac{82}{15}^{(2)}$ \\
 &  &  & 5 & $\tfrac{79}{15}^{(2)}$ \\
\midrule
$Z_3$ & 8 & 3 & 0 & $\tfrac{2}{5},\ \tfrac{8}{5},\ \tfrac{12}{5},\ \tfrac{14}{5},\ \tfrac{18}{5}^{(2)},\ 4,\ \tfrac{22}{5}^{(5)},\ \tfrac{24}{5}^{(2)},\ \tfrac{28}{5}^{(13)},\ 6$ \\
 &  &  & 1 & $\tfrac{13}{5},\ \tfrac{17}{5}^{(2)},\ \tfrac{19}{5},\ \tfrac{23}{5}^{(5)},\ \tfrac{27}{5}^{(8)},\ \tfrac{29}{5}^{(5)}$ \\
 &  &  & 2 & $\tfrac{12}{5},\ \tfrac{18}{5}^{(3)},\ \tfrac{22}{5}^{(3)},\ \tfrac{24}{5}^{(3)},\ \tfrac{28}{5}^{(9)},\ 6$ \\
 &  &  & 3 & $\tfrac{17}{5}^{(2)},\ \tfrac{23}{5}^{(4)},\ \tfrac{27}{5}^{(7)},\ \tfrac{29}{5}^{(4)}$ \\
 &  &  & 4 & $\tfrac{22}{5}^{(4)},\ \tfrac{28}{5}^{(10)}$ \\
 &  &  & 5 & $\tfrac{27}{5}^{(6)}$ \\
\midrule
$VB$ & 16 & 3 & 0 & $\tfrac{8}{5},\ \tfrac{12}{5},\ \tfrac{14}{5},\ \tfrac{18}{5}^{(4)},\ \tfrac{22}{5}^{(8)},\ \tfrac{24}{5}^{(4)},\ \tfrac{28}{5}^{(25)},\ 6$ \\
 &  &  & 1 & $\tfrac{7}{5},\ \tfrac{13}{5}^{(2)},\ \tfrac{17}{5}^{(3)},\ \tfrac{19}{5}^{(2)},\ \tfrac{23}{5}^{(10)},\ 5,\ \tfrac{27}{5}^{(17)},\ \tfrac{29}{5}^{(10)}$ \\
 &  &  & 2 & $\tfrac{12}{5}^{(2)},\ \tfrac{18}{5}^{(5)},\ \tfrac{22}{5}^{(7)},\ \tfrac{24}{5}^{(5)},\ \tfrac{28}{5}^{(18)},\ 6^{(2)}$ \\
 &  &  & 3 & $\tfrac{17}{5}^{(3)},\ \tfrac{23}{5}^{(9)},\ \tfrac{27}{5}^{(13)},\ \tfrac{29}{5}^{(8)}$ \\
 &  &  & 4 & $\tfrac{22}{5}^{(7)},\ \tfrac{28}{5}^{(18)}$ \\
 &  &  & 5 & $\tfrac{27}{5}^{(12)}$ \\
\midrule
$B_3$ & 8 & 3 & 0 & $\tfrac{12}{5},\ \tfrac{18}{5}^{(2)},\ \tfrac{22}{5}^{(4)},\ \tfrac{24}{5}^{(2)},\ \tfrac{28}{5}^{(12)},\ 6$ \\
 &  &  & 1 & $\tfrac{13}{5},\ \tfrac{17}{5}^{(2)},\ \tfrac{19}{5},\ \tfrac{23}{5}^{(5)},\ \tfrac{27}{5}^{(8)},\ \tfrac{29}{5}^{(5)}$ \\
 &  &  & 2 & $\tfrac{18}{5}^{(2)},\ \tfrac{22}{5}^{(3)},\ \tfrac{24}{5}^{(2)},\ \tfrac{28}{5}^{(9)}$ \\
 &  &  & 3 & $\tfrac{17}{5}^{(2)},\ \tfrac{23}{5}^{(5)},\ \tfrac{27}{5}^{(7)},\ \tfrac{29}{5}^{(4)}$ \\
 &  &  & 4 & $\tfrac{22}{5}^{(2)},\ \tfrac{28}{5}^{(8)}$ \\
 &  &  & 5 & $\tfrac{27}{5}^{(6)}$ \\
\end{longtable}}

\section{Numerical parameters and gaps}\label{app:numerical-parameters}
In this appendix we collect the numerical parameters of the SDPB runs that produced the plots in the main text. We also report the set of gap assumptions used to isolate the theory of three decoupled 3-state Potts models.

\begin{table}[H]
  \centering\small
  \begin{tabular}{l c c}
    \toprule
    & dark region & light region \\ \midrule
    derivative order $\Lambda$ & 11 & 19 \\
    margin $\delta$ & $0.01$ & $0.1$ \\
    recursion order $r_N$ & 25 & 25 \\
    retained-pole cutoff $\kappa$ & 20 & 20 \\
    spins & \multicolumn{2}{c}{$0\leq\ell\leq40$; $49,52,60,61,80,81,100,101$} \\
    block precision (digits) & 300 & 400 \\
    SDPB precision (bits) & 768 & 1024 \\
    \bottomrule
  \end{tabular}
  \caption{Numerical parameters for the two islands of Figure~\ref{fig:ising-island}.}
  \label{tab:ising-parameters}
\end{table}

\begin{table}[H]
  \centering\small
  \begin{tabular}{l c}
    \toprule
    derivative order $\Lambda$ & 11 \\
    recursion order $r_N$ & 25 \\
    retained-pole cutoff $\kappa$ & 20 \\
    spins & $0\leq\ell\leq20$; $49,52$ \\
    block precision (digits) & 300 \\
    SDPB precision (bits) & 768 \\
    \bottomrule
  \end{tabular}
  \caption{Numerical parameters, common to Figures~\ref{fig:potts-generic}, \ref{fig:potts-single},
  \ref{fig:potts-decoupled} and~\ref{fig:potts-coupled}.}
  \label{tab:potts-parameters}
\end{table}

\begin{table}[H]
  \centering\small
  \begin{tabular}{l c c c c c}
    \toprule
    & $c = 1.001$  & $c=1.01$ & $c=8/7$ & $c=7/5$ & $c=12/5$ \\ \midrule
    derivative order $\Lambda$ & 19 & 19 & 19 & 19 & 11 \\
    recursion order $r_N$ & 26 & 30 & 30 & 30 & 20 \\
    retained-pole cutoff $\kappa$ & $30^{*}$ & \multicolumn{4}{c}{20} \\
    spins & \multicolumn{5}{c}{$0,2,\dots,40$; $52,60,80,100$} \\
    block precision (digits) & 320 & \multicolumn{4}{c}{500} \\
    SDPB precision (bits) & 1024 & 1200 & 1400 & 1400 & 1200 \\
    \bottomrule
  \end{tabular}
  \caption{Numerical parameters for Figure~\ref{fig:z2-c1p001} and the four panels of Figure~\ref{fig:z2}. ${}^{*}$: At $c=1.001$, achieving the target interpolation accuracy required increasing the retained-pole cutoff to $\kappa=30$. For some spins and gaps, the single-interval node construction
produced repeated or unordered nodes. For each affected constraint, we tried $\kappa=32,34$ in succession until the nodes were real, distinct, and increasing. The measured interpolation errors were below $10^{-30}$ on the test grids.}
  \label{tab:z2-parameters}
\end{table}

\begin{table}[H]
  \centering\footnotesize
  \renewcommand{\arraystretch}{1.12}
  \begin{tabular}{l c c c c c c c c c c}
    \toprule
    channel & $\ell=0$ & 1 & 2 & 3 & 4 & 5 & 6 & 7 & 8 & $\ell\geq9$ \\ \midrule
    $S^+$ & $\tfrac{4}{5}$ &  & $\tfrac{18}{5}$ &  & $4$ &  & $6$ &  & $8$ & $\ell$ \\
    $X^+$ & $\tfrac{4}{5}$ &  & $2$ &  & $4$ &  & $6$ &  & $8$ & $\ell$ \\
    $A^-$ &  & $\tfrac{9}{5}$ &  & $3$ &  & $5$ &  & $7$ &  & $\ell$ \\
    $XA^-$ &  & $\tfrac{13}{5}$ &  & $\tfrac{19}{5}$ &  & $5$ &  & $7$ &  & $\ell$ \\
    $AA^+$ & $\tfrac{18}{5}$ &  & $\tfrac{18}{5}$ &  & $\tfrac{24}{5}$ &  & $6$ &  & $8$ & $\ell$ \\
    $V^+$ & $\tfrac{14}{15}$ &  & $\tfrac{32}{15}$ &  & $\tfrac{62}{15}$ &  & $\tfrac{92}{15}$ &  & $\tfrac{122}{15}$ & $\ell+\tfrac{2}{15}$ \\
    $V^-$ &  & $\tfrac{29}{15}$ &  & $\tfrac{47}{15}$ &  & $\tfrac{77}{15}$ &  & $\tfrac{107}{15}$ &  & $\ell+\tfrac{2}{15}$ \\
    $XV^+$ & $\tfrac{14}{15}$ &  & $\tfrac{32}{15}$ &  & $\tfrac{62}{15}$ &  & $\tfrac{92}{15}$ &  & $\tfrac{122}{15}$ & $\ell+\tfrac{2}{15}$ \\
    $XV^-$ &  & $\tfrac{29}{15}$ &  & $\tfrac{47}{15}$ &  & $\tfrac{77}{15}$ &  & $\tfrac{107}{15}$ &  & $\ell+\tfrac{2}{15}$ \\
    $VA^+$ & $\tfrac{44}{15}$ &  & $\tfrac{44}{15}$ &  & $\tfrac{62}{15}$ &  & $\tfrac{92}{15}$ &  & $\tfrac{122}{15}$ & $\ell+\tfrac{2}{15}$ \\
    $VA^-$ &  & $\tfrac{29}{15}$ &  & $\tfrac{47}{15}$ &  & $\tfrac{77}{15}$ &  & $\tfrac{107}{15}$ &  & $\ell+\tfrac{2}{15}$ \\
    $Z^+$ & $\tfrac{16}{15}$ &  & $\tfrac{34}{15}$ &  & $\tfrac{64}{15}$ &  & $\tfrac{94}{15}$ &  & $\tfrac{124}{15}$ & $\ell+\tfrac{4}{15}$ \\
    $Z^-$ &  & $\tfrac{31}{15}$ &  & $\tfrac{49}{15}$ &  & $\tfrac{79}{15}$ &  & $\tfrac{109}{15}$ &  & $\ell+\tfrac{4}{15}$ \\
    $B^+$ & $\tfrac{22}{15}$ &  & $\tfrac{34}{15}$ &  & $\tfrac{64}{15}$ &  & $\tfrac{94}{15}$ &  & $\tfrac{124}{15}$ & $\ell+\tfrac{4}{15}$ \\
    $B^-$ &  & $\tfrac{19}{15}$ &  & $\tfrac{49}{15}$ &  & $\tfrac{79}{15}$ &  & $\tfrac{109}{15}$ &  & $\ell+\tfrac{4}{15}$ \\
    $AZ^-$ &  & $\tfrac{31}{15}$ &  & $\tfrac{49}{15}$ &  & $\tfrac{79}{15}$ &  & $\tfrac{109}{15}$ &  & $\ell+\tfrac{4}{15}$ \\
    $AB^+$ & $\tfrac{46}{15}$ &  & $\tfrac{46}{15}$ &  & $\tfrac{64}{15}$ &  & $\tfrac{94}{15}$ &  & $\tfrac{124}{15}$ & $\ell+\tfrac{4}{15}$ \\
    $Z_3^+$ & $\tfrac{2}{5}$ &  & $\tfrac{12}{5}$ &  & $\tfrac{22}{5}$ &  & $\tfrac{32}{5}$ &  & $\tfrac{42}{5}$ & $\ell+\tfrac{2}{5}$ \\
    $Z_3^-$ &  & $\tfrac{13}{5}$ &  & $\tfrac{17}{5}$ &  & $\tfrac{27}{5}$ &  & $\tfrac{37}{5}$ &  & $\ell+\tfrac{2}{5}$ \\
    $VB^+$ & $\tfrac{8}{5}$ &  & $\tfrac{12}{5}$ &  & $\tfrac{22}{5}$ &  & $\tfrac{32}{5}$ &  & $\tfrac{42}{5}$ & $\ell+\tfrac{2}{5}$ \\
    $VB^-$ &  & $\tfrac{7}{5}$ &  & $\tfrac{17}{5}$ &  & $\tfrac{27}{5}$ &  & $\tfrac{37}{5}$ &  & $\ell+\tfrac{2}{5}$ \\
    $B_3^-$ &  & $\tfrac{13}{5}$ &  & $\tfrac{17}{5}$ &  & $\tfrac{27}{5}$ &  & $\tfrac{37}{5}$ &  & $\ell+\tfrac{2}{5}$ \\
    \bottomrule
  \end{tabular}
  \caption{Gaps $\Delta_{\rm gap}(R^\pm,\ell)$ used in Figures~\ref{fig:potts-single} and~\ref{fig:potts-decoupled}. For
  $\ell\leq6$ each gap is the lowest primary of three decoupled Potts models in that channel and spin
  (Table~\ref{tab:potts-spectrum}), or $\max(6,\ell+2n_V/15)$ if there is none with $\Delta\leq6$; for $\ell\geq7$
  it is the twist bound $\ell+2n_V/15$, with $n_V$ defined and given in Table~\ref{tab:potts-nE}. Each gap is lowered by the margin $\delta$ ($0.01$ in
  Figure~\ref{fig:potts-single}, $0.1$ in Figure~\ref{fig:potts-decoupled}), but never below $\Delta=\ell$. In
  $S^+$, $V^+$ and $Z^+$ at $\ell=0$ the gaps refer to the operators above the identity, $\phi$ and $z$.}
  \label{tab:potts-gaps-decoupled}
\end{table}

\begin{table}[H]
  \centering\small
  \renewcommand{\arraystretch}{1.15}
  \begin{tabular}{c l l}
    \toprule
    $n_V$ & irreps of $S_3\ltimes(S_3)^3$ & examples \\ \midrule
    0 & $S$, $X$, $\overline{XX}$, $A$, $XA$, $AA$, $\overline{AA}$, $AAA$, $\overline{AA}A$, $\overline{AAA}$ & $\id$, $\eps_1+\eps_2+\eps_3$, $\eps_i-\tfrac13(\eps_1+\eps_2+\eps_3)$ \\
    1 & $V$, $XV$, $VA$, $VAA$, $V\overline{AA}$ & ${\phi^a}_i$, ${\phi^a}_i(\eps_j-\eps_k)$ \\
    2 & $Z$, $B$, $AZ$, $AB$ & ${z^{ab}}_{ij}\sim{\phi^a}_i{\phi^b}_j$, ${\phi^a}_{i}\partial{\phi^b}_{j}-{\phi^b}_{j}\partial{\phi^a}_{i}$ \\
    3 & $Z_3$, $VB$, $B_3$ & ${\phi^1}_i{\phi^2}_j{\phi^3}_k$ \\
    \bottomrule
  \end{tabular}
\caption{Value of $n_V$, the number of factors of $V$, for each irrep of $S_3\ltimes(S_3)^3$ (Appendix~\ref{app:spectrum}). Irreps such as $B$, $VB$ and $Z_3$ are shorthand for $\overline{VV}$, $V\overline{VV}$ and $VVV$, respectively (see Section~2.3 of~\cite{Kousvos:2024dlz}). Each factor $V$ has $h,\bar h\geq\tfrac1{15}$, so every operator of a given irrep has twist
$\Delta-\ell\geq2n_V/15$.}
  \label{tab:potts-nE}
\end{table}

{
\bibliographystyle{JHEP.bst}
\bibliography{refs}
}

\end{document}